\documentclass{aa}  

\usepackage{graphicx}
\usepackage{txfonts}
\usepackage{comment}
\usepackage{threeparttable}
\usepackage[colorlinks,allcolors=blue]{hyperref}

\newcommand{\wha}          {$W_{\rm H\alpha}$}

\newcommand{\kmpers}      {\mbox{\rm km~s$^{-1}$}}
\newcommand{\kms} {\kmpers}

\newcommand{\iedge}{iEDGE}
\newcommand{\questna}{\emph{QueStNA}}

\defcitealias{kalinova2021}{K21}
\defcitealias{colombo2020}{C20}

\begin{document} 
    
   \title{The EDGE-CALIFA survey: Star formation relationships \\for galaxies at different stages of their evolution}

   \subtitle{}

   \author{D. Colombo\inst{1,2}\thanks{dcolombo@uni-bonn.de},
          V. Kalinova\inst{2},
          Z. Bazzi\inst{1},
          S.F. Sanchez\inst{3,4},
          A. D. Bolatto\inst{5},
          T. Wong\inst{6},
          V. Villanueva\inst{5,7},\\
          E. Rosolowsky\inst{8},
          A. Wei\ss\inst{2},
          K. D. French\inst{6},
          A. Leroy\inst{9},
          J. Barrera-Ballesteros\inst{3},
          Y. Garay-Solis\inst{3},\\ 
          F. Bigiel\inst{1},
          A. Tripathi\inst{6},
          B. Rodriguez\inst{1}
          }

   \institute{Argelander-Institut f\"ur Astronomie, University of Bonn, Auf dem H\"ugel 71, 53121 Bonn, Germany
   \and
   Max-Planck-Institut f\"ur Radioastronomie, Auf dem H\"ugel 69, 53121 Bonn, Germany
   \and
   Universidad Nacional Aut\'onoma de M\'exico, Instituto de Astronom\'\i a, AP 106, Ensenada 22800, BC, M\'exico
   \and
   Instituto de Astrof\'\i sica de Canarias, V\'\i a L\'actea s/n, 38205, La Laguna, Tenerife, Spain
   \and
   Department of Astronomy, University of Maryland, College Park, MD 20742, USA
   \and
   Department of Astronomy, University of Illinois, Urbana, IL 61801, USA
   \and
   Departamento de Astronom{\'i}a, Universidad de Concepci{\'o}n, Barrio Universitario, Concepci{\'o}n, Chile
   \and
   Department of Physics, University of Alberta, 4-181 CCIS, Edmonton, AB T6G 2E1, Canada
   \and
   Department of Astronomy, The Ohio State University, 140 West 18$^{\rm th}$ Avenue, Columbus, OH 43210, USA}

   \date{Received XXX; accepted XXX}
 
  \abstract
    {Galaxy evolution is largely driven by star formation activity or by the cessation of it, also called star formation quenching. In this paper, we present fundamental star formation scaling relations for groups of galaxies at different evolutionary stages. To do so, we used the integrated Extragalactic Database for Galaxy Evolution (iEDGE), which collects homogenised CO, optical continuum, and emission line information for 643 galaxies drawn from the CALIFA IFU dataset. By considering the patterns described by star-forming and retired regions across the galactic disc, we grouped the galaxies into different quenching stages using the emission line classification scheme, \questna. We observed that the molecular gas mass ($M_{\rm mol}$) decreases from star-forming to retired systems and so does the molecular-to-stellar mass ratio ($f_{\rm mol}$). In contrast, star formation efficiency (SFE) is largely constant in the quenching stages dominated by star formation and rapidly declines afterwards. Additionally, we observed that this rapid decline is more pronounced in the centre of the galaxies compared to the rest of the discs, reflecting the inside-out quenching often displayed by nearby galaxies. We also noticed that the relations between $M_{\rm mol}$ and the stellar mass ($M_*$) become increasingly shallow with the quenching stages; however, the relations between the star formation rate and $M_{\rm mol}$ steepen when moving from star-forming to retired systems. We also observed that a three-dimensional relation between star formation rate (SFR), $M_*$, and $M_{\rm mol}$ exists only for purely star-forming galaxies, while data points from other quenching groups are scattered across the parameter space. Taken together, these pieces of evidence indicate that the quenching of the galaxies cannot be explained solely by a depletion of the molecular gas and that a significant decrease in the SFE is necessary to retire the centre of the galaxies beyond the star formation green valley.}


   \titlerunning{Star formation relationships at different galaxy evolution stages}

   \authorrunning{D. Colombo, V. Kalinova, S. Sanchez et al.}

   \maketitle

\section{Introduction}\label{S:introduction}
Galaxies are defined by stars. Old stars establish the gravitational potential that determines galaxy structures such as spiral arms and bars \citep[e.g. ][]{binney_tremaine1987}. Young stars brighten up galaxies and input the energy and mechanical feedback that control the phases, the morphology, and the chemistry of the interstellar medium (ISM; \citealt{tielens2005}). Eventually, galaxies might stop creating new stars, and this occurrence also has an important impact on defining the life cycle of a galaxy. Referred to as `star formation quenching' (e.g. \citealt{faber2007}), the process is complex and might include a series of processes that are intimately interconnected.

Since stars form within the cold phase of the ISM (in particular, the molecular phase), the causes of star formation quenching need to be probed primarily in the status of the cold ISM. Mainly, a galaxy does not form stars because of a lack of molecular gas. Star formation quenching does not seem to happen in an entire galaxy at once. In most cases, it starts from the centre of the galaxy (e.g. \citealt{gonzalez_delgado2016,sanchez2018,kalinova2021}; or, due to particular phenomena, from the outskirts; e.g \citealt{thorp2022}), and the molecular gas might be missing in only some galactic regions. The gas can therefore be exhausted due to intense star formation, or it can be removed by internal or external processes. Such internal processes include stellar feedback, where winds from supernovae or HII regions eject the gas and results in chimneys and fountains \citep[e.g. ][]{biernacki2018,colling2018}. Active galactic nuclei (AGNs) could expel the gas that is channelled through the centre of the galaxies by stellar bars. Many studies consider AGN feedback to be the main process that quenches star formation in galaxies \citep[e.g. ][]{werner2019,piotrowska2022}, especially those based on simulation works (e.g. \citealt{dubois2016,weinberger2017,terrazas2020}). In addition, AGNs are supposed to prevent ionised gas from cooling down and rejuvenating the galaxies. Additionally, massive ($\sim10^{12}\,$M$_{\odot}$) dark matter halos can shock the gas accreted from the intergalactic medium and heat it to the virial temperature of the halo, making it unable to form new stars (e.g. \citealt{Birnboim2003}). External processes are related to the environment in which galaxies live and impact galaxies that are found in dense environments, such as clusters, groups, or merger pairs (\citealt{Peng2010},\citealt{Gunn1972, Abadi1999}). Gas can be removed by ram pressure stripping (\citealt{Larson1980, Balogh2000}). The partial (the outer envelope) or total (for low-mass galaxies) cold gas budget can also be removed during the travel of a galaxy towards a cluster centre (quenching by strangulation, e.g. \citealt{peng2015}). 

In addition to gas removal, galaxies can possess a significant amount of cold gas that is unable to efficiently form new stars (again, globally or in some regions of the galaxies). This inability can be due to the AGN or stellar feedback that transfers energy to the surrounding medium, which heats up the gas and prevents star formation (see \citealt{Husemann2018} and references therein). More often, however, these inefficiencies are attributed to galactic dynamics. Indeed, it has been observed that bulge-dominated galaxies are some of the most inefficient star-forming systems \citep[e.g.][]{bluck2014}. Bulge growth is considered the main mechanism that stabilises the gaseous disc, prevents fragmentation, and reduces star formation in the `morphological quenching' scenario (e.g. \citealt{martig2009}). Galactic centres have shown both higher and lower efficiencies in converting the gas into stars. Indeed, galactic centres can undergo periodic episodes of star formation and quenching. Those episodes are triggered by local starbursts that consume all available gas that can be brought towards the centre, for instance, by bar dynamics (e.g. \citealt{Krumholz2015}). Additionally, dynamics due to gravitational perturbances (such as streaming motions and shears) appear to play a role in altering the star formation efficiency in galaxies (in the so-called quenching via `dynamical suppression scenario', e.g. \citealt{Gensior2020}). However, these effects might appear significant only in galactic regions strongly dominated by dynamics (such as M51, see \citealt{meidt13}), as the star formation efficiency between spiral arms and inter-arm regions seems identical in normal spiral galaxies \citep{querejeta2021}. Additionally, dynamical effects can become dominant only once a considerable amount of gas is removed from the galaxies \citep{gensior2021}.

It is thus an open question as to whether quenching is driven primarily by the absence of molecular gas or a reduced star formation efficiency in that gas. Observationally, these two mechanisms are assessed through the variation in the molecular gas fraction ($f_{\rm mol}=M_{\rm mol}/M_*$, where $M_{\rm mol}$ is the molecular gas mass and $M_*$ is the stellar mass) or a reduction in the star formation efficiency (SFE=SFR/$M_{\rm mol}$, where SFR is the star formation rate). Generally, as galaxies move away from the `star formation main sequence' (SFMS; populated by star-forming galaxies; \citealt{brinchmann2004,daddi2007}) and towards the `red sequence' (populated by old retired systems), the molecular gas fraction decreases \citep{saintonge2017,bolatto2017,colombo2020}, and the low gas fraction in galaxies below the main sequence has often been invoked to explain their quenched star formation \citep[e.g. ][]{genzel2015,saintonge2016,lin2017,tacconi2018}. Nevertheless, studies based on kiloparsec-resolved datasets have shown that both decreased $f_{\rm mol}$ and SFE can contribute to bringing galaxies to quiescence \citep{piotrowska2020, wylezalek2022, villanueva2024, pan2024} and that one effect might be dominant compared to the other in different galactic regions. Indeed, the reduced SFE seems to have a dominant effect on quenching galactic centres compared to reduced $f_{\rm mol}$ \citep{colombo2020,pan2024}, while both effects contribute to retiring discs (\citealt{pan2024}; see also \citealt{barrera-ballesteros2025}).

\begin{figure*}
    \centering
    \includegraphics[width = 0.77 \paperwidth, keepaspectratio]{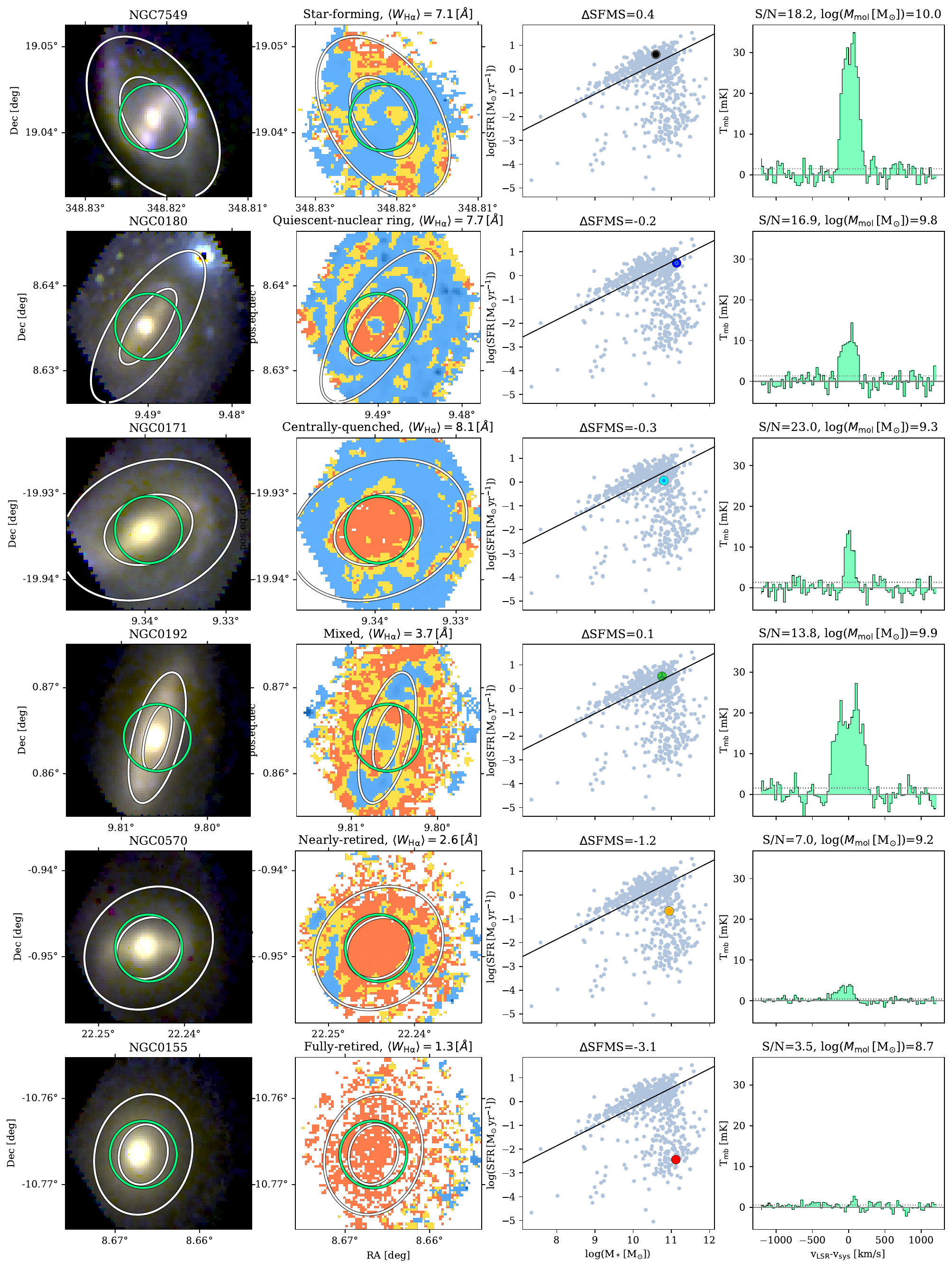}
    \caption{\emph{From top to bottom:} Examples of galaxies at different `quenching stages'. \emph{First column:} Continuum RGB images extracted from the CALIFA datacubes using $u-$ (blue), $g-$ (green), and $r-$ (red) bands. In the title, the galaxy name is given. \emph{Second column:} Discrete $W_{\rm H\alpha}$ maps. Here, blue indicates the region dominated by star formation, yellow shows diffuse gas regions, and red displays quenched regions (see text for more details). The title shows the `quenching stage' and the median $W_{\rm H\alpha}$ across the full map. In the first and second column images, white ellipses mark the locations of 1 and 2\,$R_{\rm e}$, the green circle indicates the extend and position of the APEX beam at 230\,GHz (26.2"), from which the spectra on the rightmost column are obtained. \emph{Third column:} Position of the galaxy on the SFR-$M_*$ diagram. The grey dots show the location of every galaxy in our sample. The SFMS fit (black line) is taken from \cite{cano_diaz2016}. The title gives the global distance from the SFMS, $\Delta$SFMS. \emph{Fourth column}: APEX $^{12}$CO(2-1) spectra (green) in $T_{\rm mb}$ units. All spectra have been shifted to 0\,\kms\ considering the systemic velocity of the galaxy ($V_{\rm sys}$). The dotted line represents the observation $\sigma_{\rm RMS}$. In the title, the $S/N$ of the observation is given, together with the molecular gas mass. The assembled database contains CO information for several galaxies across each quenching stage and having different distances from the SFMS.}
    \label{F:qs_example}
\end{figure*}

The emission line classification proposed by \citet[][hereafter \citetalias{kalinova2021}]{kalinova2021} represents an attempt to group galaxies by emphasizing their star formation quenching patterns rather than structural properties, as is done in morphological classification \citep{hubble1936, de_vaucouleurs1959}. The scope of this classification is to establish an observational basis for discriminating between quenching mechanisms. The classification scheme was originally defined by \citetalias{kalinova2021} on 238 galaxies observed by the Calar Alto Legacy Integral Field Area (CALIFA) integral field unit (IFU) survey \citep{sanchez2012, sanchez2016}, which provided optical emission line maps collected into pseudo-datacubes. Those datacubes have been obtained using the PIPE3D package \citep{sanchez2016b}. The classification follows the values of the equivalent width of H$\alpha$ (\wha) spaxel maps. It assumes two hard thresholds ($W_{\rm H\alpha}=3~\mathrm{\AA}$ and $W_{\rm H\alpha}=6~\mathrm{\AA}$) that are used to identify the gas as being ionised by different mechanisms. Star-forming or HII regions show $W_{\rm H\alpha}>6~\mathrm{\AA}$, finding $W_{\rm H\alpha}<3~{\mathrm{\AA}}$ indicates that the gas in the region is ionised by the old stellar population, and values of $3<W_{\rm H\alpha}<6\,\mathrm{\AA}$ are also inconsistent with star formation and could be attributed to a mix of ionisation processes \citep{CidFernandes2010,lacerda2018,sanchez2020}.

Based on these thresholds, \citetalias{kalinova2021} recognised six dominant patterns, which correspond to six different classes defined as `quenching stages' (see Fig.~\ref{F:qs_example} and \citetalias{kalinova2021} for full details). The stages are as follows: objects dominated by star formation (i.e. having $W_{\rm H\alpha}>6~\mathrm{\AA}$ across most of the galactic disc), which are referred to as `star forming' (SF); galaxies that show a quiescent nuclear ring-like structure in their centre, labelled as `quiescent-nuclear-ring' (QnR); objects retired with $W_{\rm H\alpha}<3~\mathrm{\AA}$ within $0.5\,R_{\rm e}$, called `centrally quiescent' (cQ), where $R_{\rm e}$ is the effective radius of the galaxies; galaxies that do not show a well defined pattern and are dominated by an inhomogeneous mix of ionisation effects, thus referred to as `mixed' (MX); objects that are dominated by quenching but still show a few star formation regions within $2\,R_{\rm e}$, labelled as `nearly retired' (nR); and galaxies completely retired with $W_{\rm H\alpha}<3~\mathrm{\AA}$ everywhere within $2\,R_{\rm e}$, thus `fully retired' (fR). Tentatively, the quenching stages can be further grouped into classes that resemble the classic separation of the colour-magnitude diagram as `blue cloud', `green valley', and `red sequence' (see \citetalias{kalinova2021}, their Fig.~11). The blue cloud galaxies would include all galaxies in the SF, cQ, and QnR stages. All of those stages are characterised by a global $\left< W_{\rm H\alpha}\right>\geq6\,\AA$. The green valley galaxies can be equivalent to the MX quenching stage, having a global $3\leq \left<W_{\rm H\alpha}\right><6\,\AA$. The red sequence galaxies would contain the nR and fR galaxies that show a global $\left<W_{\rm H\alpha}\right><3\,\AA$.

In addition, this emission line classification considers the `nuclear activity' of the galaxies by assuming constraints from three Baldwin-Phillips-Terlevich (BPT; \citealt{baldwin1981}) diagnostic diagrams (and their corresponding maps) that involve the [OIII], [SII], [OI], and [NII] line ratios with respect to the H$\alpha$ and H$\beta$ lines. In this respect, the classification distinguishes between active (with measurable evidence in optical lines of an AGN) and non-active (without measurable evidence in optical lines of an AGN) galaxies if at least three spaxels within $0.5\,R_{\rm e}$ populate the Seyfert region of at least two diagrams (defined by the \citealt{Kauffmann2003} separations). The active galaxies are further separated into weak AGN (wAGN) hosts if the Seyfert region spaxels show W$_{\rm H\alpha}$ values between 3 {\AA} and 6 {\AA}, and strong AGN hosts (sAGN) if $W_{\rm H\alpha}>6~\mathrm{\AA}$ (see also \citealt{Sarzi2010}, \citealt{CidFernandes2010}, \citealt{Singh2013}, \citealt{Sanchez2014}, \citealt{lacerda2018}, \citealt{lacerda2020}).

The quenching stages and the nuclear activity of the galaxies represent the two `dimensions' of the \citetalias{kalinova2021}]{kalinova2021} emission line classification called \questna. Each quenching stage (except cQ and fR by definition) can contain non-active, wAGN, or sAGN galaxies.

Studying quenching across well-defined quenching stages provides a more physically motivated framework compared to relying solely on a galaxy’s offset from the SFMS. While the SFMS primarily describes an empirical relationship between SFR and stellar mass, it does not distinguish between galaxies undergoing different quenching pathways. Two galaxies at the same distance below the SFMS may have vastly different evolutionary histories—one may be gradually exhausting its gas supply, while another experiences rapid suppression due to AGN feedback or environmental effects. For instance, the cQ and QnR galaxies in \cite{kalinova2021} show very similar distributions in the SFR-$M_*$ diagram; however, the QnR $W_{\rm H\alpha}$ pattern can be consistent with the presence of AGNs, while those of cQs cannot be. Additionally, galaxies in the QnR stage are dominated by barred galaxies (especially its nuclear active fraction) compared to the cQ stage. A quenching-stage classification explicitly tracks a galaxy’s transition from star forming to fully retired and recognises quenching as a gradual multi-phase process rather than a simple bifurcation between active and passive states. Moreover, the SFMS is a global property that averages out spatial variations in SFR, whereas a quenching-stage framework—particularly when defined using spatially resolved observations—better captures how quenching progresses inside-out, outside-in, or stochastically. This approach also enables a more direct link to the physical mechanisms driving quenching, whether internal (e.g. AGN activity, morphological transformation) or external (e.g. ram-pressure stripping, strangulation), and it aligns well with multi-wavelength observations that probe the molecular gas reservoir, stellar populations, and AGN influence. Finally, a stage-based classification is more robust regarding identification of transitioning galaxies, thus allowing for the detection of episodic star formation and rejuvenation events that a simple SFMS-based approach might overlook. By systematically tracing quenching across different stages instead of using SFMS distance as a proxy, a more comprehensive understanding of the evolutionary processes shaping galaxy populations can be gained.
While the quenching-stage classification offers a more physically motivated perspective on galaxy evolution, scaling relations between SFR, $M_*$, and $M_{\rm mol}$ remain essential for studying global statistical trends and ensuring continuity with previous works. For example, scaling relations across quenching stages could have significant implications for theoretical models of galaxy evolution. In simulations and analytical frameworks, knowing the quenching stage of a galaxy, combined with the appropriate scaling relations that we derive here, might enable more accurate estimations of molecular gas content. Since galaxies with a similar SFR or $M_*$ can be at different quenching stages and each can exhibit distinct relationships between SFR, $M_*$, and $M_{\rm mol}$, a stage-dependent approach can provide a refined method for constraining gas content in theoretical models. Therefore, by integrating the quenching-stage framework with scaling relations, this study aims to bridge the gap between observational classifications and the predictive needs of theoretical works in order to allow for a more precise understanding of how galaxies transition from star-forming to quiescent states.

In this paper, we analyse the variation of $f_{\rm mol}$ and SFE and the relationships between SFR, $M_*$, and $M_{\rm mol}$ across galaxies at different evolutionary stages in order to understand the role of the molecular gas in the emergence of the quenching patterns. For our analyses, we used data collected into the integrated Extragalactic Database for Galaxy Evolution (iEDGE; presented by \citealt{colombo2025}), which combines data from the CALIFA survey, and CO information collected by the Extragalactic Database for Galaxy Evolution (EDGE; \citealt{bolatto2017}) collaboration, which uses several millimetre-wavelength telescopes to produce a homogenised dataset that includes measurements for 643 galaxies in the local Universe. The contents of this database and the measurements used in this paper are summarised in Section~\ref{S:data}.
We classify the galaxies based on the patterns described by the distributions of star-forming and retired regions using the \questna\ classification scheme (Section~\ref{S:introduction}). In Section~\ref{S:results}, we present the results of our investigations. In particular, we analyse specific star formation (sSFR=SFR/$M_*$), $M_{\rm mol}$, SFE, and $f_{\rm mol}$ across the quenching stages defined by \questna\ (Section~\ref{SS:results_distributions}; the variation of those properties with respect to star-forming galaxies, in the centre and across the full extent of the objects (Section~\ref{SS:results_flaring}); and the scaling relations defined by those properties (Section~\ref{SS:results_relations}-\ref{SS:results_3d_relation}). We conclude with Sections~\ref{S:discussion} and \ref{S:summary}, where we present a discussion and a summary of our work. In Appendix~\ref{A:auto_class}, we describe our method to automatically classify galaxies through the \questna\ scheme. Across the paper, we assume a cosmology of $H_0 = 71\,$km$\,$s$^{-1}\,$Mpc$^{-1}$, $\Omega_{\rm m}$ =0.27,  and $\Omega_{\rm \Lambda}$ =0.73.

\section{Data}\label{S:data}
\begin{figure}
    \centering
    \includegraphics[width = 0.4 \paperwidth, keepaspectratio]{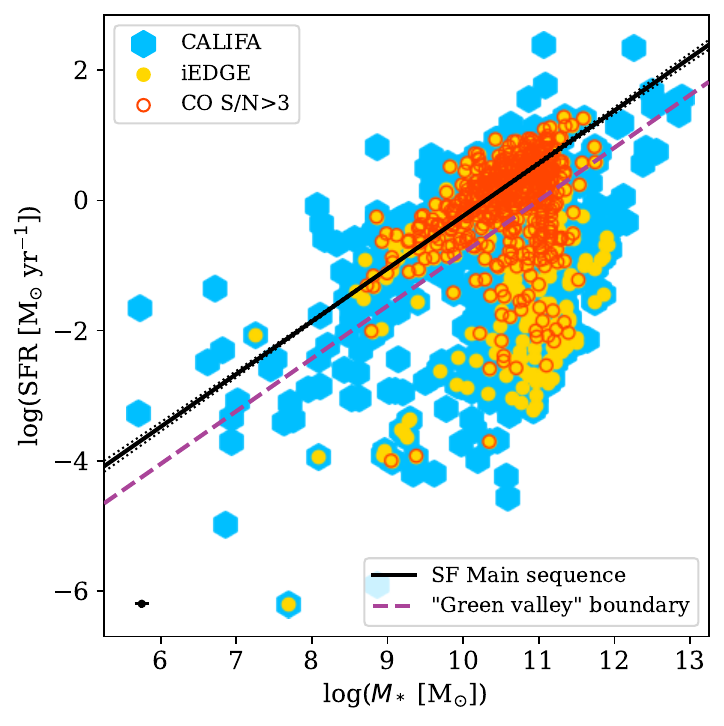}
    \caption{Diagram of SFR-$M_*$ showcasing the coverage of iEDGE compared to CALIFA. The full CALIFA sample is shown with blue hexagons, while yellow circles indicate galaxies included in iEDGE. Additionally, red circle marks CO-detected (with S/N>3) objects in iEDGE. The solid black line in both panels represents the SFMS model from \cite{cano_diaz2016}, with dotted lines showing uncertainties. The dashed purple line marks the green valley boundary from \cite{colombo2020}, positioned at $3\sigma$ (0.20 dex, following \citealt{cano_diaz2016}) below the SFMS.}
    \label{F:iedge_sfr_mstar}
\end{figure}

For our analyses, we use data collected from the \iedge\ (presented by \citealt{colombo2025}), which combines optical measurements from CALIFA, and CO information from the Combined Array for Research in Millimeter-wave Astronomy (CARMA), the Atacama large mm/sub-mm Compact Array (ACA), and the Atacama Pathfinder Experiment (APEX) telescopes (as part of the CARMA-EDGE, \citealt{bolatto2017}; ACA-EDGE, \citealt{villanueva2024}; and APEX-EDGE, \citealt{colombo2020}; surveys, respectively) to produce a dataset that includes integrated measurements for 643 local galaxies (e.g. with redshit <0.08). Galaxies observed in CO by CARMA-EDGE and ACA-EDGE predominantly populate the SFMS, whereas those in APEX-EDGE match well the CALIFA coverage of the green valley and red sequence. Indeed, CARMA-EDGE focuses on infrared-bright targets, while APEX-EDGE imposed no selection beyond the requirement that galaxies be observable with APEX (i.e., declination $<30^{\circ}$). Altogether, the CO data compiled in iEDGE provide comprehensive coverage of CALIFA galaxies with $\log(M_*/\mathrm{M}_{\odot})>8.5$ (see Fig.~\ref{F:iedge_sfr_mstar}). As the database unifies data collected from a heterogeneous set of telescopes (a single-dish radio telescope, two interferometers and an integral field spectrograph), those data have been homogenised. The homogenisation techniques include tapering of IFU data, Gaussian smoothing for interferometric data, and aperture correction of single-dish data. These techniques have produced a database containing two kinds of measurements: `global', integrated across the full galaxy extends (typically 2\,$R_{\rm e}$) and `beam', integrated across the region with the size of the APEX beam at 230\,GHz (FWHM=26.3", corresponding to a median of 1\,$R_{\rm e}$).

Specifically, from the \iedge\ in this paper, we will use median $W_{\rm H\alpha}$, integrated SFR and $M_*$ calculated from CALIFA data, and $M_{\rm mol}$ from APEX, CARMA and ACA. To summarise, SFR maps are obtained by applying the Balmer decrement method on H$\alpha$ maps, while $M_*$, inferred by the single stellar population (SSP) fit of the stellar continuum, are already included in the PIPE3D pseudo-cube provided by CALIFA. Global SFR and $M_*$ are calculated integrating across the spaxels of the maps, while beam quantities are measured from the same maps filtered by a tapering function ($W_{\rm T}$) that reproduces a 2D Gaussian map with the FWHM equal to the APEX beam at 230\,GHz (e.g. FWHM=26.3").

Molecular gas masses are obtained from CO(1-0) (for CARMA data) and from CO(2-1) (for APEX and ACA data) luminosities inferred from CO fluxes following equation 3 of \cite{solomon1997}. Global CO fluxes for CARMA and ACA are obtained from their respective databases presented by \cite{bolatto2017} and \cite{villanueva2024}, while global CO fluxes for APEX are inferred imposing an aperture correction on the CO integrated luminosity derived considering the tight correlation between CO the 12\,$\mu$m luminosity (see e.g. \citealt{gao2019,chown2021,leroy2021}). Instead, beam CO fluxes for APEX are directly given by the observation, while for CARMA and ACA, those are obtained from the central pixel spectrum of the datacubes, preventively smoothed to a 26.3" spatial resolution and a 30\,\kms\ spectral resolution, in order to mimic APEX observations. To convert from CO(1-0) luminosities to molecular gas masses we considered a CO-to-H$_2$ conversion factor ($\alpha_{\rm CO(1-0)}$) prescription that takes into account the gas-phase metallicity and the stellar mass surface density (e.g. equation 31 in \citealt{bolatto2013}). Gas-phase metallicity maps are derived from CALIFA [OIII] and [NII] emission line maps considering the O3N2 method from \cite{pettini_pagel2004}, in star formation-dominated regions, and using a radial metallicity map inferred from the corresponding mass-metallicity relation fitted by \cite{sanchez2019} to recover old stellar population dominated regions. Stellar mass surface density maps are included in the PIPE3D pseudo-cube directly provided by CALIFA. Additionally, to convert from CO(2-1) to CO(1-0) fluxes for APEX and ACA data, we built CO(2-1)-to-CO(1-0) ratio maps ($R_{21}$) that consider the tight relationship between $R_{21}$ and the SFR surface density from H$\alpha$ presented by \cite{den_brok2023}. As for the other CALIFA-derived quantities, global $\alpha_{\rm CO(1-0)}$ and $R_{21}$ are obtained as medians across the maps, while the respective beam quantities are calculated as weighted medians, where the weights are given by the tapering function $W_{\rm T}$. 
In addition, we consider the ratio between SFR and stellar mass, the specific star formation rate (sSFR=SFR/$M_*$), the ratio between SFR and molecular gas mass, the star formation efficiency (SFE=SFR/$M_{\rm mol}$ or its inverse, the depletion time $\tau_{\rm dep}$), and the ratio between molecular gas mass and the stellar mass, the molecular gas fraction ($f_{\rm mol}=M_{\rm mol}/M_*$).
Full details of the quantity derivations, models, and assumptions used to build \iedge\ are given in \cite{colombo2025}.
Throughout the paper, we will consistently use the global quantities unless indicated otherwise (with a sub- or superscript `G' for global measurements and `B' for beam measurements).

\section{Results}\label{S:results}

\subsection{Quenching stage and nuclear activity statistics}\label{SSS:sample_stats_questna}

\begin{table}
\label{T:qs_na_count}
\caption{QueStNA classification for galaxies in iEDGE.}
\centering
\begin{tabular}{c|ccc|c}
\hline
\hline
Group & Non-active & sAGN & wAGN & Total \\
\hline
SF & 276(232) & 13(11) & 2(2) & 291(245) \\
QnR & 20(18) & 4(4) & 9(9) & 33(31) \\
cQ & 59(47) & 0(0) & 0(0) & 59(47) \\
MX & 86(58) & 15(15) & 7(7) & 108(80) \\
nR & 70(38) & 3(2) & 6(6) & 79(46) \\
fR & 73(20) & 0(0) & 0(0) & 73(20) \\
\hline
Total & 584(413) & 35(32) & 24(24) & 643(469) \\
\hline
\hline
\end{tabular}
\tablefoot{Number of galaxies classified at a given `quenching stage' (left column) and showing a given `nuclear activity' (first row). Values within brackets indicate the number of CO detections (S/N>3) in a certain category.}
\end{table}

\begin{figure}
    \centering
    \includegraphics[width = 0.4 \paperwidth, keepaspectratio]{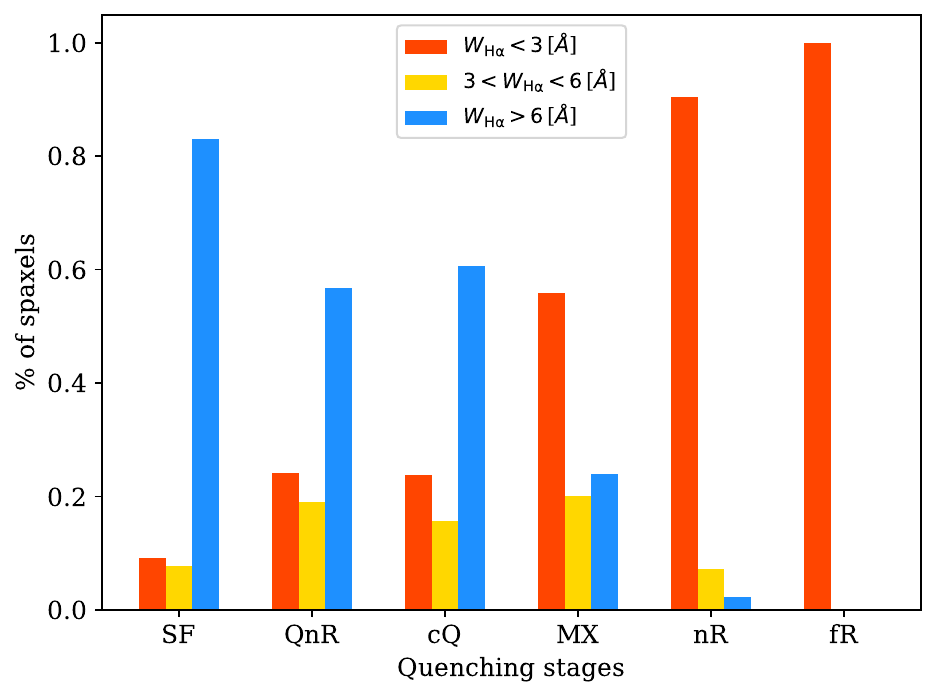}
    \caption{Percentage of spaxels dominated by star formation (blue), by diffuse gas (yellow), or quenched (red) across non-active galaxies at a given quenching stage. The number of spaxels is calculated with 2\,$R_{\rm e}$ where \questna\ classification is defined. Crossing the green valley (roughly spanned by the MX group), retired regions become dominant over the star-forming regions.}
    \label{F:sfm_whafrac}
\end{figure}

\begin{figure*}
    \centering
    \includegraphics[width = 0.9 \paperwidth, keepaspectratio]{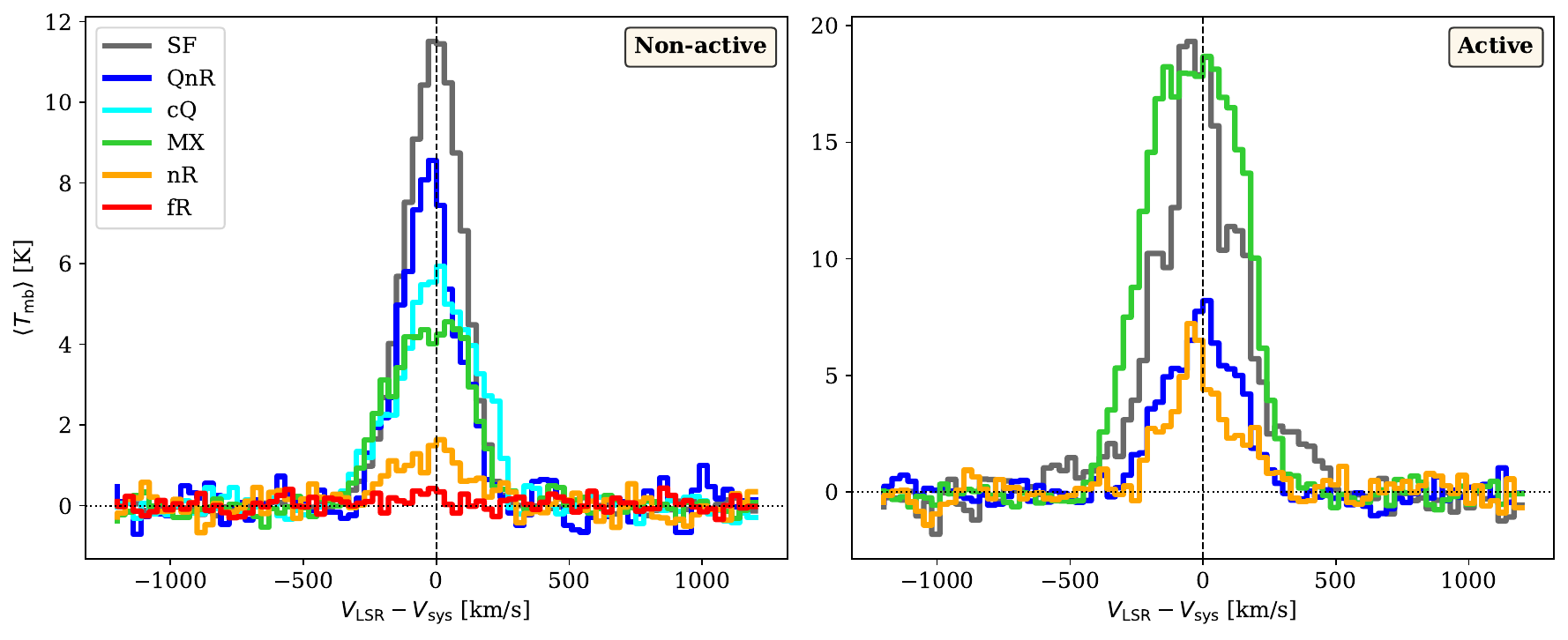}
    \caption{Spectra of CO(2-1) from APEX observations only stacked by quenching stages. In the left panel, only non-active galaxies are considered, while in the right panel, only active (wAGN and sAGN) galaxies are used. The peak of the spectra decreases from SF to fR galaxies. However, from the point of view of the stacked spectra, active but quenching galaxies with an MX or nR $W_{\rm H\alpha}$ morphology appear as molecular gas-rich as galaxy groups close to the SFMS (SF and QnR).}
    \label{F:qs_stack_spectra}
\end{figure*}

To classify the galaxy sample in the 6 proposed quenching stages and their relative nuclear activity, we automated the procedure described in \citetalias{kalinova2021}]{kalinova2021}. The classes attributed to the automatic algorithm are afterwards verified visually to generate the final classification. The details of the automatic algorithm are described in Appendix~\ref{A:auto_class}, together with the differences observed with respect to the visual classification. The numbers of galaxies classified at a given quenching stage and nuclear activity are collected in Table~\ref{T:qs_na_count}. Generally, almost half of the sample has been classified as SF. This is not surprising since most of the original CARMA galaxies are infrared-bright and, therefore, star forming in nature. Fewer objects appear to be retired in their centre either fully (cQ) or showing a quiescent-ring structure (QnR), while more galaxies show a complex gas ionisation distribution (MX stage). Finally, less than a quarter of the sampled galaxies are dominated by ionisation from the old stellar population (almost equally separated into nR and fR objects). AGN-host galaxies are generally rare in our sample and constitute less than 10\% of it (similarly to the fractions found by \citealt{lacerda2020,sanchez2018,kalinova2021,osorio-clavijo2023}). In particular, we observe a slightly higher fraction of sAGN-host compared to wAGN-host. Generally, the highest number of AGN hosts is observed within the MX stage, followed by the SF, QnR and nR stages. A representative discrete \wha\ map for galaxies on each quenching stage in our sample is shown in Fig.~\ref{F:qs_example}, together with their position in the SFR-$M_*$ diagram and their $^{12}$CO(2-1) spectrum from APEX observations. The fraction of the galaxies in each quenching stage changes slightly by considering only the CO-detected objects (with S/N>3). Given this, it is clear that SF galaxies dominate the detections. We also noted that AGN-host galaxies are all detected. 

The relative fraction of spaxels dominated by star formation, mixed effects, and the old stellar population across all galaxies in a given category is illustrated in Fig.~\ref{F:sfm_whafrac}. For the purpose of this analysis, only non-active galaxies have been used, as the $W_{\rm H\alpha}$ in spaxels where the ionisation is dominated by the AGN is not indicative of star-forming or retired regions. It is interesting to note how the fraction of star-forming and mixed spaxels decreases across the quenching stages while the fraction of retired spaxels increases. cQ and QnR classes show an almost identical fraction distribution despite the different morphological distribution of the star-forming, mixed, and retired regions. In general, while galaxies increase their distance from the SFMS, not only does the ratio between the number of star-forming and retired regions decrease, but the spatial distribution of those regions also changes.

It is also instructive to discuss now the CO spectra stacked (e.g. averaged) by the quenching stage (Fig.~\ref{F:qs_stack_spectra}). CO spectra are the basic data products provided by a millimetre-wavelength radio telescope, and their appearance is not driven by models and assumptions. For this analysis, we selected only APEX data, including also non-detected galaxies. It is interesting to notice that the amplitude of the spectra is inversely proportional to the average logarithmic distance of the galaxies at the given quenching stages (see \citealt{kalinova2021} and also the following Fig.~\ref{F:violins} and \ref{F:sfr_mstar}). Stacked spectra appear to have a largely similar FWHM, possibly because our galaxies span rather similar sets of masses. Non-significant differences are observed by using detected galaxies only. Instead, the stacked spectra for active galaxies show a different behaviour compared to their non-active counterparts at the same quenching stages. In particular, the spectra for galaxies in advanced stages of quenching (such as the MX and nR) remain comparable in amplitude to those of groups closer to the SFMS (the SF and QnR, respectively). This suggests that AGN hosts retain significant amounts of molecular gas, even in their central regions. Indeed, it is well established that active galaxies often exhibit higher gas content than non-active objects of the same stellar mass \citep[e.g.,][]{saintonge2017,koss2021}, likely due to gas inflows that fuel black hole accretion, causing the AGN events. However, the presence of such substantial gas reservoirs in the centres of these galaxies appears inconsistent with a scenario in which AGN feedback efficiently removes substantial gas amounts and directly drives quenching. We return to these aspects in Section~\ref{SS:discussion_evol}. However, a proper analysis of the properties of active and non-active galaxies at the same quenching stages and the effect of AGN feedback of quenching using iEDGE is presented elsewhere \citep{bazzi2025}. This means that the more the galaxy is quenched, the less CO bright it is. We test whether star formation quenching of the galaxies depends on the amount of molecular gas or other factors.

\subsection{Star formation property distributions}\label{SS:results_distributions}

\begin{figure}
    \centering
    \includegraphics[width = 0.4 \paperwidth, keepaspectratio]{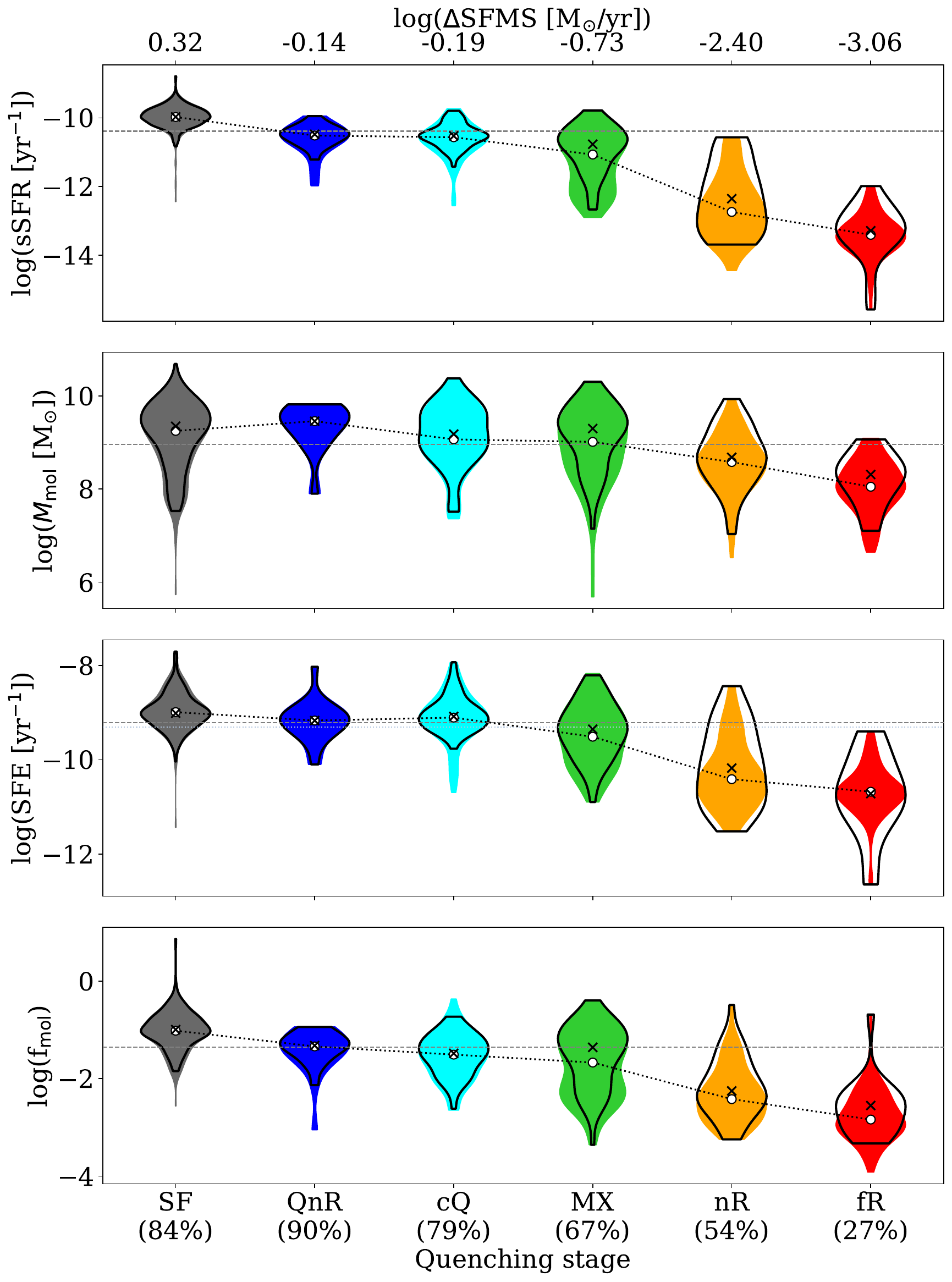}
    \caption{Violin plot representations of sSFR, $M_{\rm mol}$, SFE, and $f_{\rm mol}$ of the observed \iedge\ galaxies grouped into quenching stages. Horizontal dashed lines show the sample medians, while the white circle indicates the medians for each quenching stage. Transparent violins show the distributions of the properties, including only CO detection (S/N>3), and the respective median is indicated with a black cross. In addition, classes are ordered considering the median $\Delta$SFMS. At the bottom, the percentage of CO detections within a given class is shown. Generally, property distributions from detections and CO measurement upper limits trace well the distributions from CO-detected targets only. While SFE, on average, appears quite constant across the first quenching stages (SF, QnR, and cQ) to drop drastically afterward, $f_{\rm mol}$ shows a continued decrease from the SF to the fR class.}
    \label{F:violins}
\end{figure}

\begin{figure*}
    \centering
    \includegraphics[width = 0.9\paperwidth, keepaspectratio]{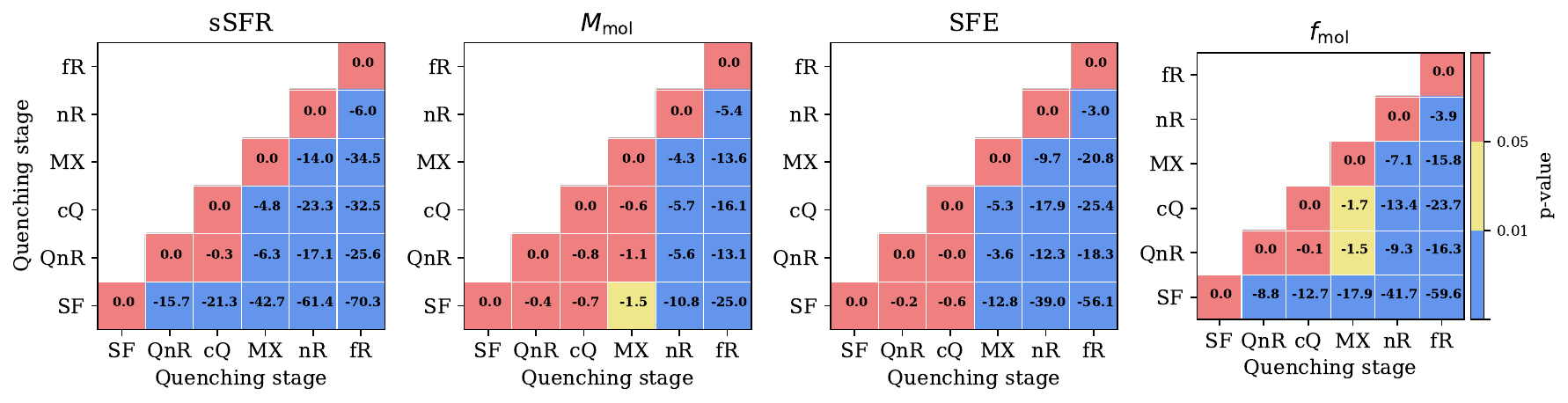}
    \caption{P-value matrices from the two-sided Kolmogorov–Smirnov test for star formation fundamental quantities from galaxies at different quenching stages. Global properties considered here are (from left to right): specific star formation rate (sSFR), molecular gas mass ($M_{\rm mol}$), star formation efficiency (SFE), and molecular gas fraction ($f_{\rm mol}$). P-value matrices are presented with discrete values: p-value$>0.05$, red; $0.01<$p-value$<0.05$, yellow; p-value$<0.01$, cyan. The $\log_{10}$(p)-values from the property distributions of a pair of quenching stages are given in logarithmic scale in the matrix cells. While the galaxies in the different quenching stages describe well-separated sSFR distributions, the SF to MX stages show similar $M_{\rm mol}$ distributions. This is then reflected in the distributions of SFE and $f_{\rm mol}$ across the quenching stages.}
    \label{F:pvalmat}
\end{figure*}

\begin{table*}
\label{T:QS_ratios}
\caption{Property averages across quenching stages.}
\centering
\renewcommand{\arraystretch}{1.8}
\begin{tabular}{c|cc|cc|cc|cc}
\hline
Group & \multicolumn{2}{|c}{$\log$(sSFR\,[yr$^{-1}$])} & 
\multicolumn{2}{|c}{$\log$($M_{\rm mol}$\,[M$_{\odot}$])} &
\multicolumn{2}{|c}{$\log$(SFE\,[yr$^{-1}$])} & \multicolumn{2}{|c}{$\log(f_{\rm mol}$)} \\
\hline
 & All & S/N$>3$ & All & S/N$>3$ & All & S/N$>3$ & All & S/N$>3$ \\
\hline
SF & $-9.97^{+0.20}_{-0.19}$ & $-9.97^{+0.19}_{-0.18}$ & $9.25^{+0.40}_{-0.65}$ & $9.35^{+0.35}_{-0.54}$ & $-8.99^{+0.28}_{-0.22}$ & $-9.01^{+0.23}_{-0.19}$ & $-1.02^{+0.25}_{-0.24}$ & $-0.99^{+0.24}_{-0.16}$ \\
QnR & $-10.51^{+0.14}_{-0.30}$ & $-10.47^{+0.12}_{-0.16}$ & $9.46^{+0.25}_{-0.31}$ & $9.46^{+0.25}_{-0.22}$ & $-9.17^{+0.16}_{-0.21}$ & $-9.17^{+0.13}_{-0.16}$ & $-1.34^{+0.15}_{-0.23}$ & $-1.32^{+0.15}_{-0.10}$ \\
cQ & $-10.56^{+0.16}_{-0.28}$ & $-10.52^{+0.15}_{-0.13}$ & $9.07^{+0.54}_{-0.30}$ & $9.18^{+0.46}_{-0.30}$ & $-9.11^{+0.24}_{-0.23}$ & $-9.08^{+0.19}_{-0.20}$ & $-1.51^{+0.30}_{-0.43}$ & $-1.48^{+0.29}_{-0.24}$ \\
MX & $-11.07^{+0.50}_{-0.97}$ & $-10.77^{+0.36}_{-0.51}$ & $9.01^{+0.47}_{-0.61}$ & $9.30^{+0.35}_{-0.58}$ & $-9.51^{+0.42}_{-0.54}$ & $-9.35^{+0.42}_{-0.40}$ & $-1.67^{+0.45}_{-0.60}$ & $-1.36^{+0.36}_{-0.64}$ \\
nR & $-12.74^{+0.64}_{-0.59}$ & $-12.35^{+0.72}_{-0.87}$ & $8.58^{+0.29}_{-0.38}$ & $8.68^{+0.48}_{-0.43}$ & $-10.41^{+0.48}_{-0.39}$ & $-10.18^{+0.75}_{-0.64}$ & $-2.42^{+0.35}_{-0.35}$ & $-2.25^{+0.49}_{-0.30}$ \\
fR & $-13.40^{+0.28}_{-0.30}$ & $-13.29^{+0.48}_{-0.23}$ & $8.05^{+0.42}_{-0.32}$ & $8.31^{+0.23}_{-0.36}$ & $-10.68^{+0.27}_{-0.24}$ & $-10.72^{+0.52}_{-0.38}$ & $-2.84^{+0.40}_{-0.25}$ & $-2.55^{+0.15}_{-0.47}$ \\
\hline
\hline
\end{tabular}
\tablefoot{Median with the scatter giving the interval between the median and the 25$^{\rm th}$ percentile (-) or between the 75$^{\rm th}$ percentile and the median (+) for the sSFR, $M_{\rm mol}$, SFE, and $f_{\rm mol}$ distributions. Those quantities are shown for all galaxies and the detected fraction (with S/N>3) at a given quenching stage.}
\end{table*}

The property distributions of sSFR, $M_{\rm mol}$, SFE, and $f_{\rm mol}$ for galaxies at the same quenching stage are shown in Fig.~\ref{F:violins} using violin plot representation. For this analysis and the rest of the paper, we will consider only non-active galaxies, e.g. 585 targets (407 detected). The study of the star formation properties of the active versus non-active galaxies will be presented elsewhere \citep{bazzi2025}. The violin distributions corresponding to the different quenching stages in the panels are sorted out based on the sample median distance from the SFMS ($\Delta$SFMS), where we considered the fit of \cite{cano_diaz2016} of the SFMS (see their table~1). First of all, we can immediately observe that in most cases, distributions for the full sample (that include both detections and upper limits for the CO fluxes) are almost equivalent to the distributions drawn from detections only. Therefore, considering only sources with detections, even for the classes where a small fraction of galaxies have been detected, will not change our conclusions significantly. Distributions of sSFR of stages closer to the SFMS (e.g. SF, QnR, and cQ) overlap a little with the later stages representing the most retired galaxies (nR and fR). The sSFR distribution for the mixed galaxies (e.g. the green valley galaxies) instead span a more extended parameter space. The trend of the averages is clearly descending across the quenching stages.  

Instead, the distributions of $M_{\rm mol}$ largely overlap, almost independently from the quenching stage. Some differences are observed, though, in particular regarding fR galaxies that do not contain the largest values of molecular gas mass in the full sample, and the QnR and cQ galaxies where, on the opposite, the lower values are not observed. This trend is also reflected in the behaviour of the median across the stages: it is slightly above the sample median ($\sim1.7\times10^9$\,M$_{\odot}$) for the SF and QnR, similar to the sample median for cQ and MX groups, and a factor 3-11 below for the retired groups. We note that by considering only detected targets, the median $M_{\rm mol}$ of the nR class is similar to the sample median.

The distributions of SFE are clearly $>10^{-11}$\,yr$^{-1}$ from SFMS galaxies up to green valley galaxies (the MX group), while they drop drastically to values as low as $10^{-12}$\,yr$^{-1}$ for the retired groups. Nevertheless, all categories (except fR) show larger SFE values above $10^{-9}$\,yr$^{-1}$. It is also interesting to note that, on average, SFE does not significantly change up to green valley galaxies, and it remains close to the sample median of $\sim 4\times10^{-10}$\,yr$^{-1}$. Again, the most significant drop is observed for the retired groups, where the SFE decreases by 20 to 30 times for the nR and fR stages, respectively. 

In general, the trends of SFE and $f_{\rm mol}$ distributions across the quenching stages are similar. For $f_{\rm mol}$ we find a significant drop in the distributions to lower values for the retired groups, which is surprisingly similar to SFE behaviour: $f_{\rm mol}$ medians of nR and fR groups are 20 to 35 times lower than the sample median ($\sim6.7\times10^{-2}$). Interestingly, some of the galaxies in the MX group showed some of the highest values of $f_{\rm mol}$ in the sample, similar to the SF group. The $f_{\rm mol}$ distribution of MX is the only one that shows a clear bi-modality, which might be largely driven by the upper limits of $M_{\rm mol}$. Additionally, the median of the SF group is slightly ($\sim2\times$) larger than the sample median. 

The preliminary conclusion at this point is that both decrements of SFE and $f_{\rm mol}$ both contribute to bringing galaxies to quiescence. Even if galaxies start to be quenched in their centre, their global properties do not change significantly until the star formation is halted across the full galactic disc. Green valley galaxies (the MX group) do not only possess a well-defined quenching pattern but span properties that are common to both star-forming and quenching types. Notably, some nearly retired targets also show overlaps with the distributions of star-forming targets.

It is now interesting to study whether the distributions for different star formation-related properties across quenching stages can be drawn for the same parental distribution. This test will also help to further assess the ability of the \questna\ classification to separate galaxies into groups with specific characteristics. To do so, we considered the p-value obtained by the two-sample Kolmogorov-Smirnov (KS) test\footnote{\url{https://docs.scipy.org/doc/scipy/reference/generated/scipy.stats.ks_2samp.html}} and represented it using discrete p-value matrices (as in Fig.~\ref{F:pvalmat}). Significant differences between the distributions will typically give p-values of either $<0.05$ or $<0.01$. We divided the analysis on the full sample of galaxies and the detected galaxies only. In terms of sSFR, the distributions appear significantly different everywhere. The only exception is given by the cQ and QnR classes, where the p-value is slightly above 0.05. This is expected as quenching stages appear to follow closely the evolution of the galaxies across the SFR-$M_*$ diagram. Considering the other properties, in general, the nR and fR group distributions appear to be the most different compared to each other and the other categories, showing very low p-values almost everywhere. Instead, $M_{\rm mol}$ distributions for galaxy groups above the retired sequence appear statistically similar. The most significant statistical difference is given by $f_{\rm mol}$ and the SF group compared to the others (e.g. cQ, QnR, and MX). Indeed, as observed in Fig.~\ref{F:violins}, SF galaxies appear to possess the highest amount of molecular gas mass (compared with their stellar mass) with respect to the other quenching stages.

\subsection{Variation of star-forming properties}\label{SS:results_flaring}
As described in Section~\ref{S:data}, the intrinsic nature of our data, which consists on one hand of kiloparsec-resolved CO and optical maps, and on the other hand by single-pointing CO spectra, allowed us to build a database formed by global and beam properties, where beam properties are measured, essentially, within 1\,$R_{\rm eff}$ (see \citealt{colombo2025}).

\begin{figure*}
    \centering
    \includegraphics[width = 0.85 \paperwidth, keepaspectratio]{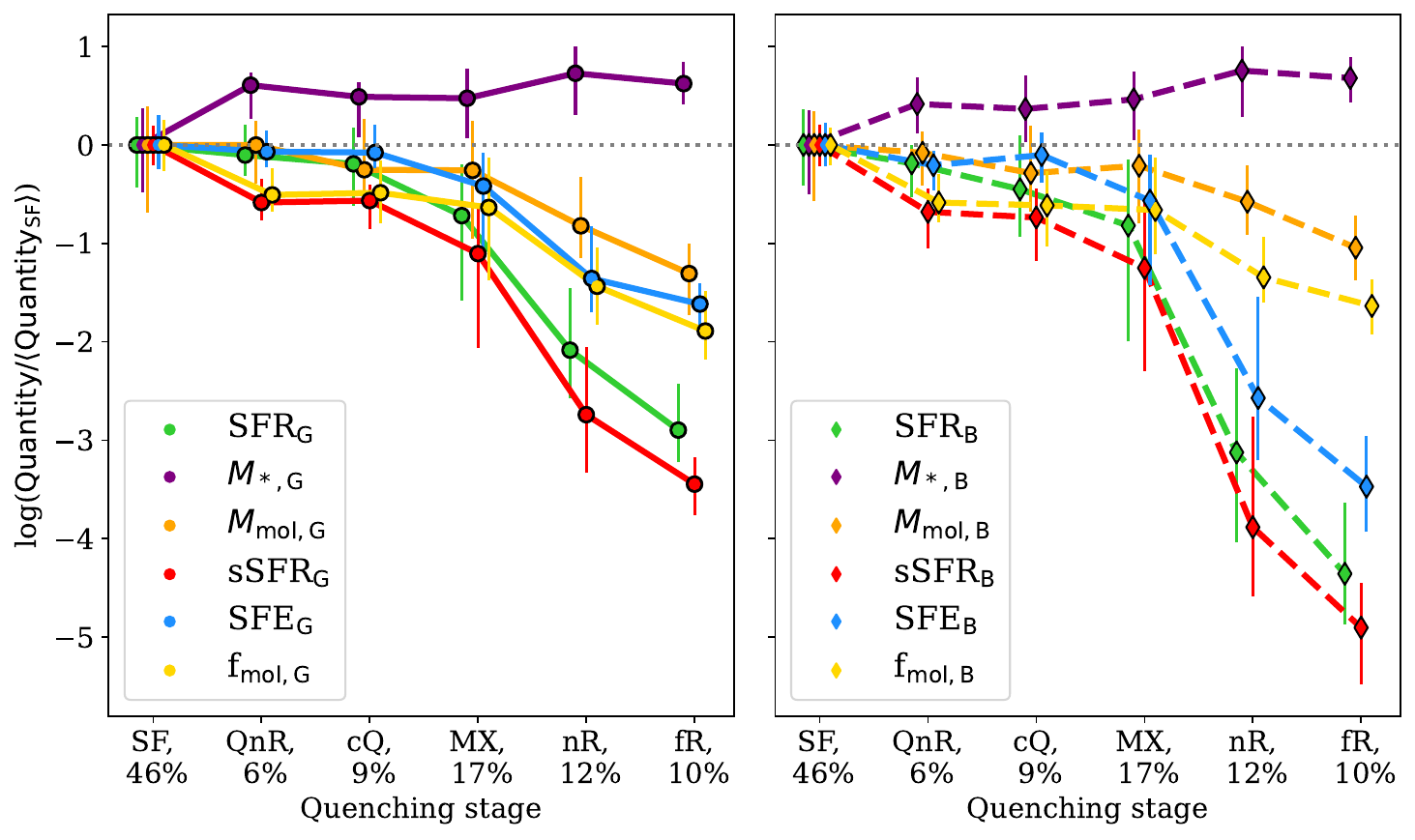}
    \caption{`Flaring' plots (see text for additional details) of several quantities related to the star formation process (star formation rate, SFR, green; stellar mass, $M_*$, purple; molecular gas mass, $M_{\rm mol}$, orange; specific star formation rate, sSFR, red; star formation efficiency, SFR, cyan; molecular gas fraction, $f_{\rm mol}$, yellow). In the left panel, `global' quantities are considered, and in the right panel, `beam' quantities. For these analyses, we calculated the median (circles for global quantities and diamonds for beam quantities) and interquartile range (coloured bars) of the distribution of a given quantity and for a given quenching stage divided by the quantity median from the SF group. Full (global quantities) and dashed (beam quantities) are drawn to guide the eye across the quenching stages. Additionally, we used only CO-detected galaxies (S/N>3), and the percentage of the galaxies in a given quenching stage compared to the total sample used in this analysis is indicated in the x-axis labels. All properties (except sSFR) appear to decrease, on average, from the SF to the fR stages. However, the decrement of SFR (and relative quantities) is more significant (especially in the centre of the galaxies).}
    \label{F:flaring}
\end{figure*}

In this section, we extended the analysis presented in Section~\ref{SS:results_distributions} by observing how the star formation-related properties of the various quenching stages vary compared to the SF group. To avoid systemic biases introduced by $M_{\rm mol}$ upper limits, we will use only CO-detected galaxies. Additionally, we studied how the properties vary globally and in the inner region of the galaxies (Fig.~\ref{F:flaring} left and middle panels, respectively). To do so, we adopted the `flaring plot' technique developed in a different context \citep[see e.g.][]{leroy2017,den_brok2022}, where, for a given property, we show the logarithm of the median of a given quenching stage distribution divided by the median of the SF class distribution. Indeed, as the galaxies become more quenched, the points defined by each star formation-related property spread and the plot `flares'. 

Firstly, galaxies in each quenching stage are, on average, more massive than SF targets. The stellar mass behaviour is quite flat, meaning that it does not vary much from the QnR to the fR galaxies. The variations are fundamentally identical for beam and global stellar masses. The molecular gas mass variation with respect to the SF stage becomes more significant for the nR and fR stages. The variation in the inner galaxy largely reflects the global variation of $M_{\rm mol}$. The SFR follows a similar behaviour, but the variation with respect to the SF stage starts to become more important at the MX stage. Interestingly, the variation of SFR in the inner galaxy for the nR and fR stages is more important than the global variation of this property: approximately one order of magnitude lower from the MX to the retired targets (globally) and two orders of magnitude lower in the inner regions. Given that the trend of the $M_*$ from the entire galaxies and their inner regions is similar, the behaviour of sSFR mirrors the SFR behaviour, with more significant decrements observed in the inner galaxy rather than globally. The variation of the $f_{\rm mol}$ follows the variation of $M_{\rm mol}$ across the quenching stages and for the inner regions and the entire galaxies. Still, we observed that, on average, $f_{\rm mol}$ is half of an order of magnitude lower for QnR, cQ, and MX types with respect to the SF group. This variation decreases to 1-1.5 orders of magnitude for the nR and fR stages. Global and beam $f_{\rm mol}$ variations closely follow each other (see Fig.~\ref{F:flaring}, right panel). The SFE displays the most interesting behaviour. On average SFE for QnR and cQ targets is similar to SF objects. The SFE starts to decrease in the MX group and drops significantly (of one order of magnitude) for the nR and fR groups. However, while the global variation of SFE is fundamentally identical to the variation (global and beam) of $f_{\rm mol}$, the drop of SFE for the retired groups is way more remarkable in the inner galactic regions (more than 2 orders of magnitude with respect to the SF stage). This evidence suggests that both the stellar and molecular gas mass distributions remain relatively uniform across galactic discs, regardless of quenching stage. In contrast, the central regions of galaxies in advanced quenching stages exhibit significant drops in SFR compared to their outer discs. However, these conclusions are based on our beam and global measurements, which provide only indirect constraints on the distribution of gas, stars, and SFR in the outer regions. To robustly confirm these findings, a statistically significant sample of kiloparsec-resolved CO observations of galaxies across the most advanced quenching stages is needed.

\subsection{Fundamental scaling relations for star formation}\label{SS:results_relations}
Having established some crucial links between the quenching stages of the galaxies, the star formation-associated properties, and the property-related ratios, we now focus our attention on the scaling relations between those properties, in order to understand if correlations exist and how significant they are for each quenching stage.

\begin{figure*}
    \centering
    \includegraphics[width = 0.85 \paperwidth, keepaspectratio]{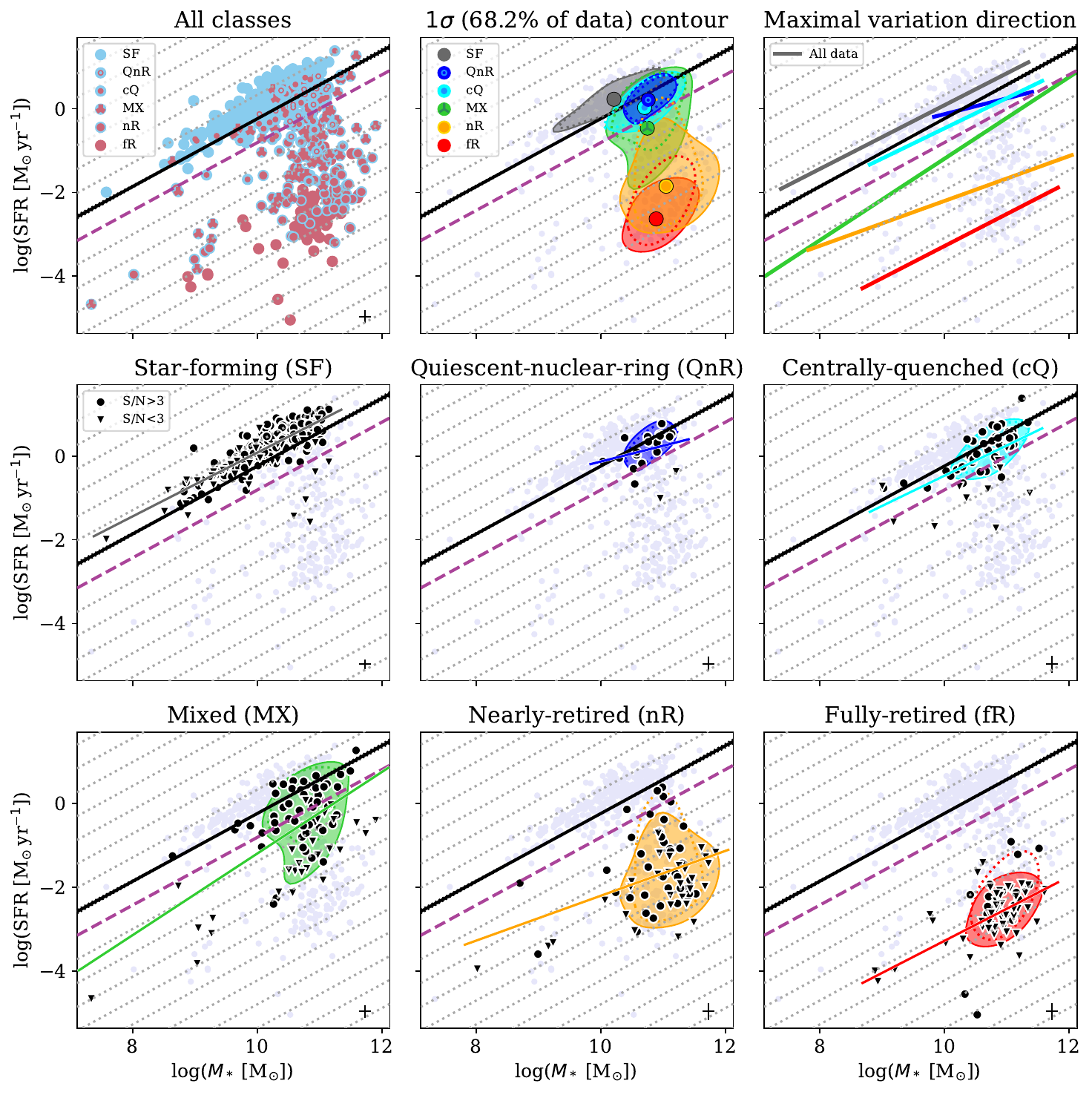}
    \caption{Diagram of SFR-$M_*$ from the \iedge\ in different representations. \emph{Upper left}: Individual data points are drawn with markers that reflect the $W_{\rm H\alpha}$ bi-dimensional distribution across the galaxy disc. Here, blue shows regions where $W_{\rm H\alpha}>6\,\AA$, while red shows regions where $W_{\rm H\alpha}<3\,\AA$. \emph{Upper middle and lower panels:} Full (dotted) contours encompass 68.2\% (1$\sigma$) of the data points in a given quenching stage for all galaxies (CO-detected galaxies only). \emph{Upper right and lower panels:} Direction defined by the Bayesian linear regression performed by \emph{linmix} for the data points in each quenching stage is shown. In the lower panels, the distribution of data in a given quenching stage is highlighted. Additionally, CO-detected galaxies are shown with circles, while triangles indicate galaxies for which an upper limit in the CO luminosity is used. Across the panels, the black solid line shows the SFMS fit from \cite{cano_diaz2016} and the purple dashed line is the green valley `boundary' defined in \cite{colombo2020}, which is located $3\sigma$ (equal to 0.20 dex as in \citealt{cano_diaz2016}) below the SFMS. Dotted lines are also spaced of $3\sigma$ from each other. Galaxies in the various quenching stages span different regions of the SFR-$M_*$ diagram.}
    \label{F:sfr_mstar}
\end{figure*}

\begin{figure*}
    \centering
    \includegraphics[width = 0.85 \paperwidth, keepaspectratio]{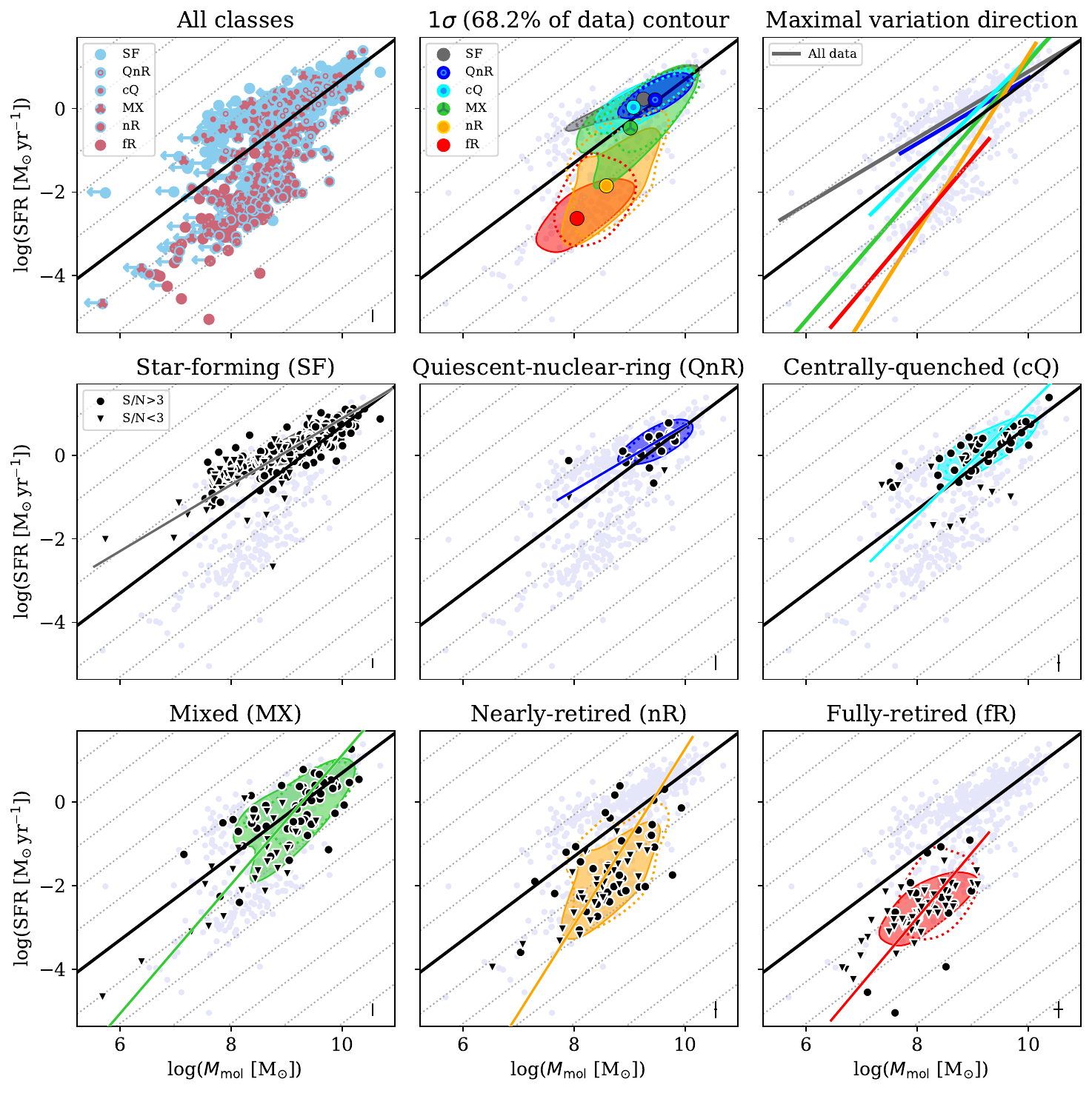}
    \caption{Diagram of SFR-$M_{\rm mol}$ from the \iedge\ in different representations. Symbols and conventions follow Fig.~\ref{F:sfr_mstar}. The black solid line across the panel shows the locus where $\tau_{\rm dep}=2\,$Gyr. Additionally, dotted lines show constant $\tau_{\rm dep}$ from (top to bottom) 10$^{-5}$\,Gyr to 10$^6$\,Gyr. When moving from quenching stages dominated by star formation to the ones dominated by retired regions, the steepness and scatter of the SFR-$M_{\rm mol}$ increases.}
    \label{F:sfr_mmol}
\end{figure*}

\begin{figure*}
    \centering
    \includegraphics[width = 0.85 \paperwidth, keepaspectratio]{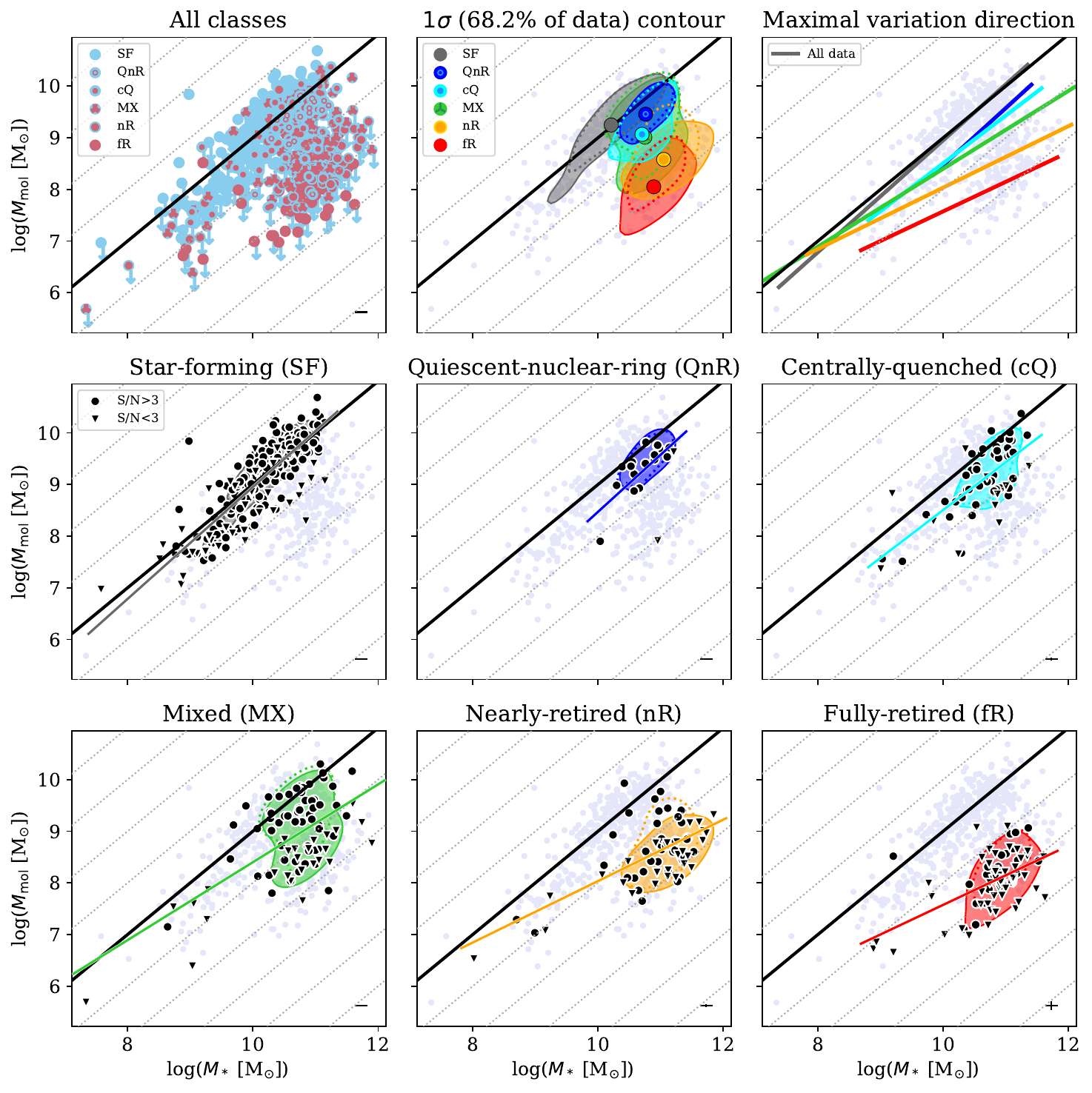}
    \caption{Diagram of $M_{\rm mol}-M_*$ from the \iedge\ in different representations. Symbols and conventions follow Fig.~\ref{F:sfr_mstar}. Across the panels, the solid black line across the panel shows the locus where $f_{\rm mol}=0.1$. Additionally, dotted lines show a constant $f_{\rm mol}$ from (bottom to top) 10$^{-5}$ to 10$^4$. Moving from quenching stages dominated by star formation to the ones dominated by retired regions, the relationships between $M_{\rm mol}-M_*$ shift to lower molecular gas fractions and become shallower.}
    \label{F:mmol_mstar}
\end{figure*}

The scaling relations between global SFR, $M_*$ and $M_{\rm mol}$ are presented in Fig.~\ref{F:sfr_mstar}, \ref{F:sfr_mmol}, and \ref{F:mmol_mstar}. We used several investigation methods in all the figures to establish correlations between the parameters. In the left panel of the upper row of each scaling relation figure, we represented the quenching stages with a marker that recalls the appearance of their typical discrete $W_{\rm H\alpha}$ maps (see Fig.~\ref{F:qs_example}). In the middle panel of the upper row, we used bi-dimensional `kernel density estimate' (KDE) contours to indicate the region of the parameter space that contains 68.2\% (e.g. 1$\sigma$) of data points in each quenching stage. In the right panel of the upper row, we showed the linear models across the quenching stages estimated using \emph{linmix} (\citealt{kelly2007}; available online\footnote{\url{https://github.com/jmeyers314/linmix}}). This algorithm uses a hierarchical Bayesian model and performs a linear regression by taking into account uncertainties on both variables. Additionally, \emph{linmix} handles non-detections by treating them as upper limits using a hierarchical Bayesian approach. Instead of assigning them fixed values, it models their likelihood distribution, ensuring they contribute to the overall fit without biasing the results. The method employs a latent variable framework, where the true values of the censored data are treated as unknowns, sampled during the MCMC process. This allows \emph{linmix} to properly account for measurement uncertainties and selection effects while estimating the underlying relationship. It also provides the intrinsic scatter of the data across the model. The list of \emph{linmix}-estimated parameters are collected in Table~\ref{T:QS_fit}, together with Pearson's ($r_{\rm P}$\footnote{\url{https://docs.scipy.org/doc/scipy/reference/generated/scipy.stats.pearsonr.html}}) and Spearman's ($r_{\rm S}$\footnote{\url{https://docs.scipy.org/doc/scipy/reference/generated/scipy.stats.spearmanr.html}}) correlation coefficients calculated with {\sc scipy}. The coefficients indicated whether two quantities are linearly correlated or monotonically growing, respectively. We calculated those coefficients for the full samples and only for the CO-detected samples. In the lower rows of the figures, the scaling relations are separated for each quenching stage. 
Additionally, we assessed whether a bi-dimensional distribution related to a given quenching stage can be drawn from the distribution of another quenching stage using the bi-dimensional version of the Kolmogorov-Smirnov test discussed in \cite{fasano_franceschini1987} and implemented in the {\sc ndtest} {\tt python} package\footnote{\url{https://github.com/syrte/ndtest}}.
\\

\begin{figure*}
    \centering
    \includegraphics[width = 0.9\paperwidth, keepaspectratio]{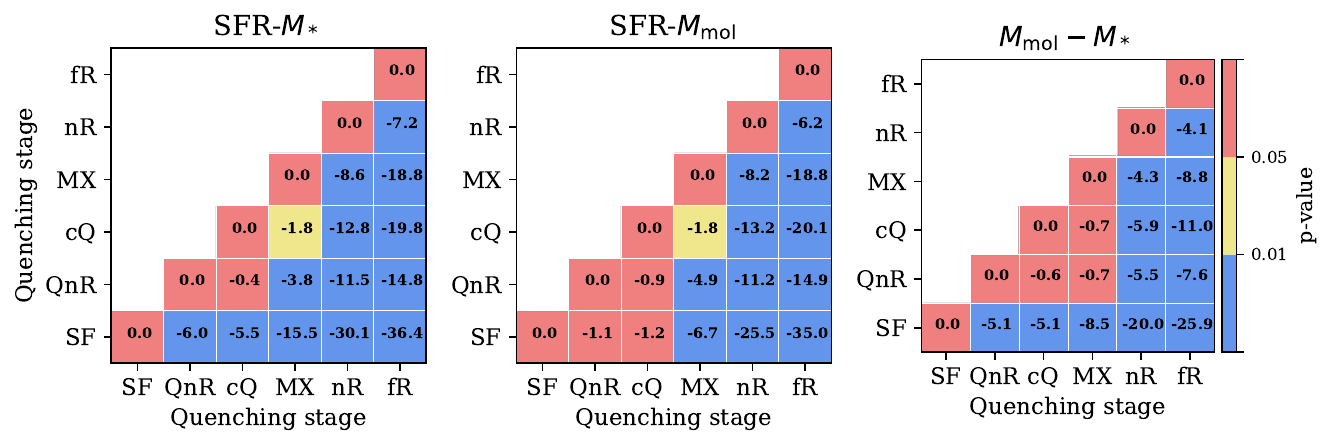}
    \caption{P-value matrices from the two-sided bi-dimensional Kolmogorov–Smirnov test for star formation-related scaling relations from galaxies at different quenching stages. Symbols and conventions follow Fig.~\ref{F:pvalmat}. Considering also the overlap of the data across the diagrams, the test shows that the bi-dimensional distribution described by SFR and $M_{\rm mol}$ of cQ and QnR galaxies can be largely drawn by the distribution of the SF class. Instead, these galaxy group bi-dimensional distributions seem to be largely discernible from the MX group distribution from $M_{\rm mol}$ and $M_*$.}
    \label{F:pvalmat_scalrels}
\end{figure*}

We begin our exploration by considering the scaling relations between the global SFR and $M_*$ presented in Fig.~\ref{F:sfr_mstar}. This diagram has already been explored across quenching stages of galaxies by \cite{kalinova2021} using a sample of 238 CALIFA systems. Here, we expand the sample. In the upper left panel of Fig.~\ref{F:sfr_mstar}, we notice how the quenching spreads from the centre to the outskirts of the galaxies already across the high-mass side of the SFMS. In the low-mass segment of the SFMS, all galaxies basically belong to the SF stage. Additionally, most of the SF galaxies appear located slightly above the SFMS, while QnR and cQ targets (which completely overlap, as confirmed by the 2D KS test, Fig.~\ref{F:pvalmat_scalrels}, left column) are located mostly below it. Additionally, the contours of these stages are encompassed within the MX contour, which spreads from the SFMS to the `green valley'. Interestingly, the green valley location is shared with (mostly detected) galaxies belonging to the nR class. Targets in the fR group are in the `retired' sequence, and their contour partially overlaps with the nR contour. Pearson's and Spearman's correlation coefficients and scatter estimated by \emph{linmix} indicate that a positive, tight linear correlation exists only for the SF class (Table~\ref{T:QS_fit}). The correlation between the cQ and fR subsamples are at most moderate (correlation coefficients up to $0.6$) and show higher scatter values, while there is basically no correlation between the SFR and $M_*$ for the galaxies in the QnR, MX, and nR groups. While this for the QnR class can be due to the small number of data points, the same is not true for MX and nR stages, which behave as `transition groups' on the SFR-$M_*$ diagram: from the SFMS to the green valley and from the green valley to the retired sequence, respectively. The contours of those classes therefore appear elongated along the y-axis instead of along the linear fit directions. The regression slopes for SF, cQ, and fR indicate that the relations for these classes are almost parallel with each other and with respect to the SFMS fit.
\\

The SFR-$M_{\rm mol}$ for global quantities equivalent to the ``Kennicutt-Schmidt's relationship'' (e.g. \citealt{schmidt1959}, \citealt{kennicutt1998}) for integrated measurements and extended to retired galaxies is shown in Fig.~\ref{F:sfr_mmol}. We can immediately notice that a sort of `main sequence' where most of the SF galaxies are located is not observed in the plot, the region of the diagram characterised by $\tau_{\rm dep}=2$\,Gyr is formed by galaxies at several quenching stages (Fig.~\ref{F:sfr_mmol}, upper left), even at the nR and fR stages. SF targets are found across the $\tau_{\rm dep}=2$\,Gyr loci, even if several objects (even detected) appear to be shifted to lower depletion times. Galaxies in the QnR and cQ classes are also generally located across the $\tau_{\rm dep}=2$\,Gyr locus. Indeed, SF, QnR, and cQ galaxy distributions can be drawn from the same parental distribution as indicated by the 2D KS test (Fig.~\ref{F:pvalmat_scalrels} middle column). From the MX, moving to the nR and fR groups, data contours (and medians) start to extend from the $\tau_{\rm dep}=2$\,Gyr loci to diagram regions characterised by lower SFR and $M_{\rm mol}$, e.g. lower SFE (or longer $\tau_{\rm dep}$). The bi-dimensional distributions related to the retired stages cannot be drawn from the same parental distribution or the distribution of star-forming targets. It is also interesting to notice that the correlation coefficients are quite high in each stage ($\sim0.7-0.9$) and the scatter is quite low (of the order of 0.2 dex), indicating that significant correlations between SFR and $M_{\rm mol}$ exist for all quenching stages (and galaxies in the SF stages are strongly correlated across this diagram). However, the relationships of the retired groups appear weaker when only detections are considered. It is interesting to notice that the slopes of the relations become steeper going from SF to fR stages. However, we observed sublinear relations for the SF and QnR groups. For the former, this might be related to the gas-poor (and possibly low-mass) galaxies in the group that tend to deviate from the $\tau_{\rm dep}=2$\,Gyr line, while gas-rich objects appear to be well described by it. These observations reveal that a sort of Kennicutt-Schmidt's law exists at each stage of galaxy evolution, but quenching galaxies appear less able to use their molecular gas budget to form new stars (hence showing low SFE).

The global $M_{\rm mol}-M_*$ relation, which has been also recently called `molecular gas main sequence' for star-forming galaxies \citep[e.g.][]{lin2019,sanchez2021} is represented in Fig.~\ref{F:mmol_mstar} with various methods and across the quenching stages. In some way, this relation reflects the behaviour of the SFR$-M_*$ relation as it is largely driven by the stellar mass: we can recognise a molecular gas main sequence described by galaxies dominated by star formation, while galaxies that start to be quenching-dominate the high-mass part of the diagram (Fig.~\ref{F:mmol_mstar}, upper row and left panel). Galaxies in the SF group are largely organised across the $f_{\rm mol}=0.1$ line, but the (strong) correlation (see Table~\ref{T:QS_fit}) defined by these targets is slightly steeper than this line. The objects that are quenched in their centre (part of the QnR and cQ classes) are mostly found with $f_{\rm mol}<0.1$. Galaxies within the MX group, though, extend from the molecular gas main sequence to regions in the diagram dominated by systems that are largely quenched (belonging to the nR and fR groups). The \emph{linmix} fits (Table~\ref{T:QS_fit}) show that the relationships become increasingly shallower from the SF group (where the relation is linear) to the fR stage (where the relation is sub-linear). Nevertheless, proper monotonic or linear correlations seem to exist only for the SF and cQ groups. For the other classes, Pearson's and Spearman's correlation coefficients indicate that the relations are weak, especially when only CO detections are considered. Additionally, the most significant scatter ($\sim0.5$ dex) is observed for the MX objects. Indeed, their distribution appears bi-modal, and the bi-modality seems to be driven by the separation between molecular gas detections and non-detections. As in the SFR-$M_*$ diagram, the MX stage acts as a transition group from the star-forming to the retired galaxies. Further, we observed that bi-dimensional distributions described by $M_{\rm mol}$ and $M_*$ can be drawn by the same parental distribution for QnR, cQ and MX, which is the similar behaviour observed for the SFR-$M_*$ diagram. This is possibly driven by the stellar mass. In the limit in which the stellar mass dominates the galactic potential, it looks plausible that the formation and the presence of molecular gas are driven by the stellar mass amount. However, this appears true only for star-forming galaxies, while for quenching and passive systems, more stars do not necessarily mean more (molecular) gas.

\begin{table*}
\label{T:QS_fit}
\caption{Scaling relation parameters across quenching stages.}
\setlength{\tabcolsep}{2pt}
\centering
\renewcommand{\arraystretch}{1.5}
\begin{tabular}{c|ccccccc}
\hline
& \multicolumn{7}{|c}{$\log$(SFR)=$q+m\log(M_{\rm *})$} \\
\hline
Group & $m$ & $q$ & $\langle x\rangle$ & $\langle y\rangle$ & Scatter & $r_{\rm P}$ & $r_{\rm S}$ \\
\hline
SF & $0.76^{+0.02}_{-0.02}$ & $-7.54^{+0.22}_{-0.21}$ & 10.21 & 0.23 & 0.06 & 0.77 & 0.76 \\
QnR & $0.38^{+0.28}_{-0.28}$ & $-3.92^{+3.04}_{-3.01}$ & 10.76 & 0.21 & 0.20 & 0.23 & 0.38 \\
cQ & $0.72^{+0.10}_{-0.10}$ & $-7.71^{+1.06}_{-1.07}$ & 10.77 & 0.02 & 0.26 & 0.61 & 0.62 \\
MX & $0.98^{+0.09}_{-0.09}$ & $-10.95^{+0.99}_{-0.97}$ & 10.75 & -0.45 & 0.68 & 0.64 & 0.40 \\
nR & $0.54^{+0.11}_{-0.11}$ & $-7.57^{+1.16}_{-1.15}$ & 11.05 & -1.82 & 0.77 & 0.39 & 0.17 \\
fR & $0.77^{+0.09}_{-0.10}$ & $-10.94^{+1.01}_{-1.01}$ & 10.9 & -2.63 & 0.32 & 0.58 & 0.46 \\
\hline
\hline
& \multicolumn{7}{|c}{$\log$(SFR)=$q+m\log(M_{\rm mol})$} \\
\hline
Group & $m$ & $q$ & $\langle x\rangle$ & $\langle y\rangle$ & Scatter & $r_{\rm P}$ & $r_{\rm S}$ \\
\hline
SF & $0.79^{+0.02}_{-0.02}$ & $-7.06^{+0.19}_{-0.20}$ & 0.23 & 9.25 & 0.13 & 0.82 (0.85) & 0.86 (0.85) \\
QnR & $0.78^{+0.12}_{-0.09}$ & $-7.03^{+0.83}_{-1.13}$ & 0.21 & 9.46 & 0.10 & 0.64 (0.52) & 0.65 (0.69) \\
cQ & $1.32^{+0.16}_{-0.13}$ & $-11.95^{+1.17}_{-1.46}$ & 0.03 & 9.07 & 0.26 & 0.66 (0.83) & 0.75 (0.84) \\
MX & $1.54^{+0.09}_{-0.08}$ & $-14.32^{+0.71}_{-0.81}$ & -0.46 & 9.01 & 0.24 & 0.80 (0.65) & 0.72 (0.66) \\
nR & $2.11^{+0.21}_{-0.17}$ & $-19.84^{+1.49}_{-1.84}$ & -1.83 & 8.6 & 0.21 & 0.66 (0.53) & 0.62 (0.49) \\
fR & $1.58^{+0.13}_{-0.11}$ & $-15.40^{+0.91}_{-1.08}$ & -2.6 & 8.06 & 0.15 & 0.69 (0.58) & 0.71 (0.43) \\
\hline
\hline
& \multicolumn{7}{|c}{$\log(M_{\rm mol}$)=$q+m\log(M_{\rm *})$} \\
\hline
Group & $m$ & $q$ & $\langle x\rangle$ & $\langle y\rangle$ & Scatter & $r_{\rm P}$ & $r_{\rm S}$ \\
\hline
SF & $1.08^{+0.03}_{-0.03}$ & $-1.88^{+0.29}_{-0.29}$ & 10.21 & 9.25 & 0.15 & 0.84 (0.86) & 0.81 (0.85) \\
QnR & $1.09^{+0.32}_{-0.31}$ & $-2.43^{+3.34}_{-3.39}$ & 10.76 & 9.46 & 0.27 & 0.52 (0.80) & 0.57 (0.72) \\
cQ & $0.92^{+0.09}_{-0.09}$ & $-0.73^{+0.95}_{-0.91}$ & 10.77 & 9.05 & 0.24 & 0.70 (0.78) & 0.62 (0.69) \\
MX & $0.75^{+0.07}_{-0.07}$ & $0.89^{+0.71}_{-0.73}$ & 10.75 & 9.01 & 0.44 & 0.63 (0.44) & 0.36 (0.34) \\
nR & $0.59^{+0.06}_{-0.06}$ & $2.09^{+0.62}_{-0.63}$ & 11.08 & 8.6 & 0.23 & 0.64 (0.51) & 0.44 (0.30) \\
fR & $0.57^{+0.07}_{-0.07}$ & $1.85^{+0.78}_{-0.79}$ & 10.89 & 8.05 & 0.23 & 0.55 (0.40) & 0.49 (0.63) \\
\hline
\hline
\end{tabular}
\tablefoot{Slope ($m$), intercept ($q$), median $x-$quantity, median $y-$quantity, intrinsic scatter, and correlation coefficient for the SFR-$M_*$, $M_{\rm mol}-M_*$, and SFR-$M_{\rm mol}$ relations for galaxies at a given `quenching stage' (as indicated in the first column). Quantities (expect $r_{\rm P}$ and $r_{\rm S}$) are obtained via the \emph{linmix} algorithm. Slope and intercept are provided as median with the uncertainties giving the interval between the median and the 25$^{\rm th}$ percentile (-) or between the 75$^{\rm th}$ percentile and the median (+). The intrinsic scatters are provided as medians of the relative posterior distributions. The Pearson ($r_{\rm P}$) and Spearman ($r_{\rm S}$) correlation coefficients for the full sample and for the sample of only CO-detected galaxies (within brackets) are indicated.}
\end{table*}

\subsection{Three-dimensional scaling relations across quenching stages}\label{SS:results_3d_relation}
\begin{figure}
    \centering
    \includegraphics[width = 0.4 \paperwidth, keepaspectratio]{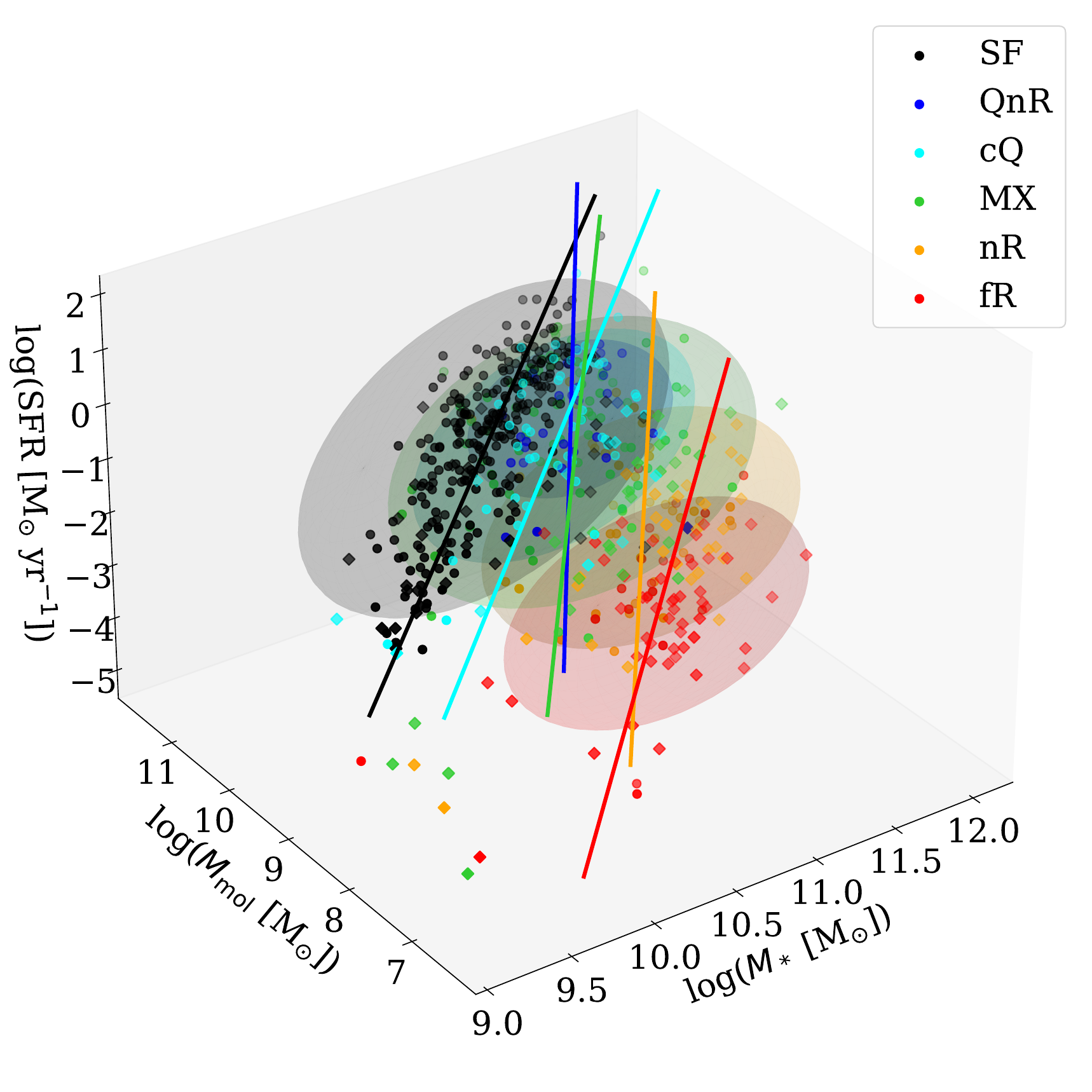}
    \caption{Three-dimensional star formation scaling relations involving star formation rate (SFR), stellar mass ($M_*$), and molecular gas mass ($M_{\rm mol}$). Different colours indicate galaxies at given quenching stages. Additionally, for each group, the confidence ellipsoids and the direction of maximal variation in the space (full lines) are calculated with PCA. Circles show CO-detected galaxies, while diamonds indicate CO non-detected galaxies (with S/N<3). Only SF galaxies form a well-defined linear distribution in the three-dimensional space.}
    \label{F:sfrels_3d}
\end{figure}

In recent years, the possibility that SFR, $M_*$, and $M_{\rm mol}$ could describe a relationship in the space has often been discussed, for both kiloparsec-resolved and integrated measurements, but generally considering SFMS galaxies \citep[e.g.][]{lin2019,sanchez2021,barrera-ballesteros2021}. Therefore, in this section, we checked whether a three-dimensional relation also exists for galaxies in other quenching stages. The relation is represented in Fig.~\ref{F:sfrels_3d} using principal component analysis (PCA) ellipsoids, where the direction of maximal variation of the data points in space is shown.  

From a visual inspection of the figure, it is clear that the relations compared to the SF galaxies move in space and become steeper in the direction of the SFR and the $M_{\rm mol}$. Here, we do not provide proper fits of the 3D relations, but we can assess whether relationships are linear in space by measuring the ratios between the eigenvalues that define the 3D ellipsoids. Eigenvalues are indicated with $\lambda$ with a number that gives them ranking: `1' for the main, `2' and `3' for the secondary ones. These ratios are collected in Table~\ref{T:3d_lambda_ratio}. The ratios show that $\lambda_1$ is 3-4$\times$ larger than $\lambda_2$ and $\lambda_3$ only for the SF group, and it is consistent when only CO detections are included. This means that there is a clear direction of maximal variation for the quantities involved in the relation of the SF class and a linear relation in space, as assessed by other studies, such as \cite{sanchez2021}.
For the other groups, the ratios are generally lower, e.g. $\lambda_1/\lambda_{2,3}\sim2-3$. In those cases, the data points are more scattered, the three-dimensional distributions are more ``spheroidal'' without a well-defined direction in space.

\begin{table}
\caption{Three-dimensional scaling relation parameters.}
\label{T:3d_lambda_ratio}
\centering
\begin{tabular}{c|cc}
\hline
Group & $\lambda_1/\lambda_2$ & $\lambda_1/\lambda_3$ \\
\hline
SF & 3.42 & 4.27 \\
QnR & 2.32 & 2.57 \\
cQ & 2.42 & 2.94 \\
MX & 2.57 & 2.96 \\
nR & 1.95 & 2.75 \\
fR & 2.27 & 2.41 \\
\hline
\hline
\end{tabular}
\tablefoot{Ratio between the eigenvalues calculated through the three-dimensional PCA on the SFR-$M_*-M_{\rm mol}$ space for each quenching stage. Eigenvalues are ordered from the major ($\lambda_1$) to the minor ($\lambda_3$).}
\end{table}

\section{Discussion}\label{S:discussion}
Throughout this study, we have demonstrated a novel approach to exploring star formation quenching by effectively combining integrated measurements (such as SFR, $M_*$, $M_{\rm mol}$; and their ratios) with spatially resolved optical emission line diagnostics (such as $W_{\rm H\alpha}$ maps). In the following, we explore the implications of our findings, highlighting the interplay between these parameters and the innovative insights they offer into the processes governing star formation quenching.

\subsection{Scaling relations across quenching stages}

In the local Universe, the SFR-$M_*$ relation has been broadly studied in both integrated \citep[see e.g. ][]{brinchmann2004,bluck2014,renzini_peng2015,leslie2020,sanchez2021,leja2022} and resolved \citep[see e.g.][]{sanchez2013,cano_diaz2016,hsieh2017,bolatto2017,lin2019,enia2020,ellison2020,sanchez2021,baker2022,wong2024} forms from star-forming to retired systems. Nevertheless, the recent study of \cite{baker2022} has shown that this relation (at least on kiloparsec scales) might be a by-product of the relationships between SFR and $M_*$ with $M_{\rm mol}$ (but see \citealt{barrera-ballesteros2021,sanchez2021}).

Based on the seminal work of \cite{kennicutt1998}, the scaling relation between SFR and $M_{\rm mol}$ has been often measured mostly across star-forming main sequence galaxies, including low- \citep[e.g. ][]{leroy2005}, high-mass \citep[e.g. ][]{rownd_young1999}, and star-burst galaxies \citep[e.g.][]{gao_solomon2004} from kiloparsec \citep[e.g. ][]{leroy2013,utomo2017} down to cloud scales \citep{kreckel2018,querejeta2021,sun2023}. Generally these studies point towards a (or close to) unitary slope for the SFR-$M_{\rm mol}$ relation \citep[e.g. ][]{murgia2002,wong_blitz2002,heyer2004,bigiel08,schruba11,leroy2013,utomo2017,dey2019, ellison2021,sun2023}, indicating a strong link between SFR and molecular gas. More recently, this relation has also been calculated for early-type systems \citep[ however, including atomic gas]{davis2014} and quenching, green valley galaxies \citep{lin2022}. Both studies concluded that the retiring systems present SFR-$M_{\rm mol}$ relations with intercepts lower than star-forming galaxies, indicating that quenching systems form galaxies with lower SFEs compared to main-sequence galaxies.

The relationship between $M_{\rm mol}$ and $M_*$ has received less attention in the literature, and it has been analysed mostly on kiloparsec-scales \citep[e.g. ][]{wong2013,bolatto2017,lin2019, barrera-ballesteros2020,barrera-ballesteros2021,morselli2020,ellison2021,sanchez2021,ellison2024}, but also at cloud-scales (e.g. $\sim100$\,pc scale; e.g. \citealt{pessa2021,pessa2022}). These studies also found that these quantities can also be described by a (close to) linear relation. As in the case of the SFR-$M_{\rm mol}$, retired regions in galaxies show similar slope relationships but shifted towards lower intercepts \citep{ellison2021}. 

While most of these studies are based on intensive quantities (e.g. surface densities), in this paper, we have measured the relationships using extensive quantities (e.g. integrated quantities). Nevertheless, as CALIFA is a diameter selected survey and its IFU covers typically 2\,$R_{\rm eff}$ in the galaxies, the extensive quantities are derived from the same galactic regions in the targets. \cite{sanchez2021}, in particular, have found integrated and kiloparsec-resolved measurements described the same star formation relations and deviations from those relations are observed mainly at sub-kiloparsec scales (see \citealt{pessa2021}; due, for example, to the influence of feedback, see e.g., \citealt{kruijssen2018,schinnerer2019,chevance2020,kim2021,zakardjian2023}). Therefore, our conclusions can be potentially applied to kiloparsec-resolved studies, too.

In Section~\ref{SS:results_relations}, we have measured the star formation relations across quenching stages. Slopes increase while intercepts decrease, moving from star-forming to retired systems for the SFR-$M_{\rm mol}$ relation. Additionally, we observed that SF targets showed a relation with a slope below unity ($\sim0.77$, possibly due to the low stellar mass and gas-poor in the SF sample) that is similar to the QnR galaxies. However, this slope more than doubles moving to green-valley galaxies (MX group) and largely retired systems (nR and fR). Additionally, scatter increases across the quenching sequence, and it reaches the highest value for MX galaxies. This evidence points towards a scenario where star formation in retired systems, although still responding to the molecular gas supply, proceeds with reduced efficiency compared to star-forming galaxies. This diminished efficiency may arise from the absence of the positive feedback loop that typically sustains star formation—where active star formation enhances the conditions for further star formation \citep[e.g.][]{ostriker2010}. In retired systems, this feedback may be disrupted, leading to conditions such as high turbulence, low gas densities in critical regions, or heating that prevents gas from cooling and condensing into stars. Alternatively, external or internal regulatory mechanisms could suppress star formation activity, even in the presence of molecular gas. This said, the behaviour described above can still be driven by the intrinsic sensitivity limitation of the surveys involved in the database, which does not allow us to access very low molecular gas masses (although, the APEX survey, which data constitute the largest part of the database, was designed to reach gas surface density $\sim2$\,M$_{\odot}$\,pc$^{-2}$, well below the HI-to-H$_2$ transition regime of 10\,M$_{\odot}$\,pc$^{-2}$). Indeed, the retired groups are largely influenced by non-detections (in particular, the fR stage is dominated by them) and the relationships we measured can be driven by upper limits (see Section~\ref{SS:results_relations}). This results in only moderately correlated SFR and $M_{\rm mol}$ values in CO-detected galaxies. Therefore, we cannot fully establish with our sample that `retired KS laws' actually exist.

The conclusions from the SFR-$M_{\rm mol}$ diagrams can be reinforced by $M_{\rm mol}-M_*$ relations across the quenching stages. While those quantities appear to be strongly and linearly correlated for the SF group, the slope of the relations seems to decrease across the quenching stages. This indicates that although galaxies tend to be slightly more massive at advanced quenching stages, the higher gravitational potential (in the limit to which it can be mainly attributed to the stellar mass) is unable to sustain a high molecular gas fraction, hence a reduced molecular gas fraction overall. Therefore, star-forming galaxies maintain a higher supply of molecular gas, which fuels star formation, while early-type galaxies gradually deplete or lose this gas, contributing to their passive evolutionary state. However, these conclusions rely on the validity of the $M_{\rm mol}-M_*$ relationships for the retired groups (nR and fR quenching stages). In these groups, both Pearson's and Spearman's correlation coefficients are relatively low (< 0.5; see Section~\ref{SS:results_relations}), and the trends are largely influenced by non-detections, introducing significant uncertainties. Additionally, a linear relation appears not appropriate to describe the data in the MX stage, where the scatter is significant.

Both \cite{lin2019} and \cite{sanchez2021} concluded that SFR, $M_{\rm mol}$, and $M_*$ describe a 3D line in the space rather than a higher dimensional shape such as a plane. \cite{sanchez2021} suggested a hidden parameter (such as the mid-plane pressure controlled by stellar feedback, e.g. \citealt{barrera-ballesteros2021,ellison2024}) could be responsible for the simultaneous existence of the three relationships. Both studies involved star-forming galaxies or star-forming spaxels. Here (Section~\ref{SS:results_3d_relation}), we extended these results using integrated quantities and including galaxies at different evolutionary stages. Through the eigenvalues that described the orientation of the relations in space, we concluded that a proper 3D relation exists only for SF galaxies. On one hand, this might be simply driven by the fact that CALIFA and our database are not complete in terms of stellar mass at each SFR range. For example, the SFMS extends to lower stellar-mass galaxies compared to the green valley and retired sequences. On the other hand, if the star formation is self-regulated by stellar feedback (e.g. \citealt{ostriker2010}), the quenching of the galaxies could imply, once again, that this controlling agent is less present, contributing to the retirement of the galaxies via the removal of the pressure support. Once again, these conclusions rely on the validity of the relationships measured for the retired groups, where non-detections dominate and introduce significant uncertainties in the interpretation. Future observations with higher sensitivity, capable of detecting the molecular gas reservoirs in these systems, will be essential to confirm and refine the trends reported here.

\subsection{The role of SFE and $f_{\rm mol}$ variations across quenching stages}

\begin{figure}
    \centering
    \includegraphics[width = 0.4 \paperwidth, keepaspectratio]{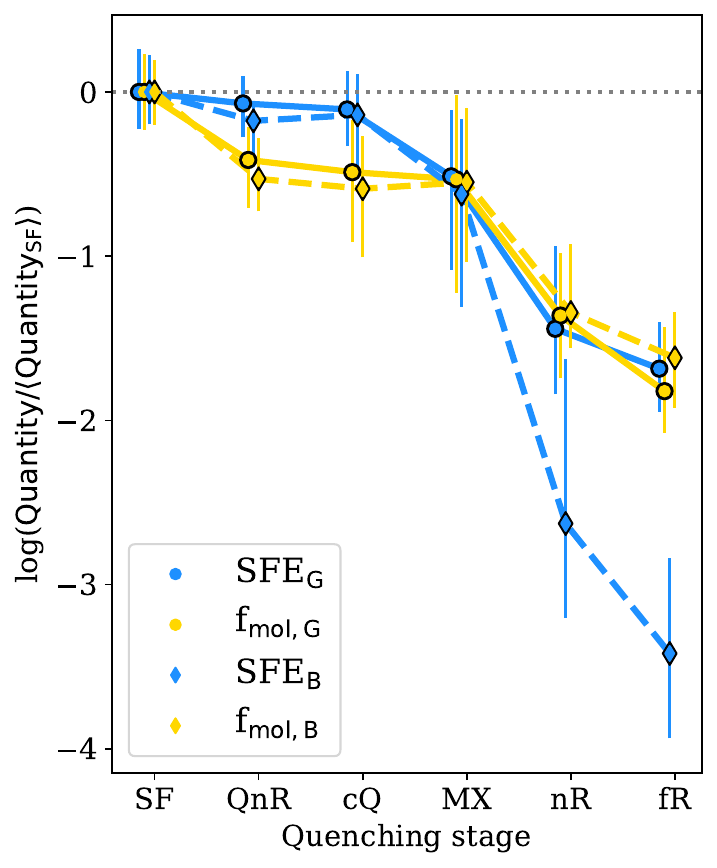}
    \caption{`Flaring' plots highlighting the behaviour of global and beam SFE (blue) and $f_{\rm mol}$ (yellow). The symbols and conventions follow Fig.~\ref{F:flaring}. While the evolution of the galaxies from the SF to the MX group is mostly driven by decrements of $f_{\rm mol}$, the significant drop (from MX to the nR/fR groups) in SFR in the centre of the galaxies is dominated by decrements of SFE.}
    \label{F:flaring_sfe_fmol}
\end{figure}

In the previous section, we observed how, despite the details on the existence and slope of the scaling relations across the quenching stages, data samples tend to move towards loci in the diagrams corresponding to progressively lower values of SFE and $f_{\rm mol}$ while quenching proceeds across the galaxies.

However, precisely understanding whether the retirement of galaxies is mainly driven by variations of available molecular gas or variations of the efficiency at which stars are formed is fundamental to discerning between the mechanisms responsible for the star formation quenching. The study of \cite{pan2024}, using 34 galaxies drawn from the ALMaQUEST sample, showed that the quenching of galaxies can be driven by a single mode (variations of SFE, $f_{\rm mol}$, or both parameters together) that act across the full galactic disc. This is observed for half of their sample. Nevertheless, the centre of the galaxies (within 0.5\,$R_{\rm eff}$) seems to be dominated by decrements of SFE. \cite{villanueva2024} with a sample of 60 CALIFA galaxies observed with ACA (the same objects included in the \iedge) observed that both SFE and $f_{\rm mol}$ decrease in the centre of green valley galaxies, concluding that both modes are necessary to explain the inside-out quenching.

Our analyses allowed us to study galaxies showing specific star formation/retirement patterns (quenching stages) and can give new insights into this issue. In Section~\ref{SS:results_distributions}, we observed that, on average, the relative amount of molecular gas, parametrised with $f_{\rm mol}$, constantly decreases from the star-forming to the quenching and the retired targets. Instead, the SFE appears largely constant for galaxies that are fully star forming or quenched in their centre. This is also observed in the scaling relations, where the contours of SF, QnR, and cQ galaxies largely overlap in the SFR-$M_{\rm mol}$ diagram. This might indicate that variations of $f_{\rm mol}$ globally drive the quenching of galaxies. Nevertheless, we could derive different conclusions by considering central variations of the two parameters with respect to global ones. To check this, in Fig.~\ref{F:flaring_sfe_fmol}, we isolated the behaviours of (global and beam) SFE and $f_{\rm mol}$ across quenching stages using the flaring plot technique.

We can observe that galaxies that are largely quenched within 0.5\,$R_{\rm eff}$ (such as the ones in the cQ and QnR categories) exhibit SFE comparable to SF objects. In addition, the global and central SFE in these objects are equivalent. This reflects the results from nearby star-forming discs, where often flat SFE behaviours across the discs are observed (see e.g. \citealt{leroy2008,utomo2017,muraoka2019,villanueva2024}). In these galaxies, the quenching seems to be driven by decrements of $f_{\rm mol}$, which appears to be lower than in SF targets. However, this might be driven by the fact that the SF category includes a larger number of low-mass objects compared to the QnR and cQ groups. This is because CALIFA does not encompass a significant number of low-mass green valley and retired systems in contrast to other IFU surveys (see \citealt{bluck2019}, their Fig.~1, for example). Among galaxies across the green valley (e.g. in the MX category), decrements of both SFE and $f_{\rm mol}$ (compared to the SF group) follow each other. Nevertheless, as observed in Section~\ref{SS:results_relations}, the MX groups span a significant fraction of parameter space in the SFR$-M_{\rm mol}$ and $M_{\rm mol}-M_*$ diagrams, and therefore these conclusions are valid `on average'. The SFE is reduced with respect to QnR and cQ galaxies, while $f_{\rm mol}$ is comparable to these categories. In this aspect, we may conclude that the quenching of the green valley galaxies in our sample is driven by both variations of SFE and $f_{\rm mol}$. However, crossing the MX group, the situation changes drastically. While global variations of SFE follow central and global variations of $f_{\rm mol}$, the SFE in the centre of the largely retired galaxies is strongly reduced with respect to the other quantities, compared to the SF group. This reduction could reflect the cessation of positive feedback mechanisms that sustain star formation, as SFE may decline further once molecular gas is depleted, leaving insufficient fuel to maintain those processes. The central quenching of galaxies in the nR and fR classes is therefore clearly driven by a reduced SFE. This observation reflects the general conclusion of \cite{colombo2020}, where after a decrement of the molecular gas availability, the centre of galaxies is rapidly quenched by significantly decreased SFE.

\subsection{The evolution of galaxies across quenching stages and quenching mechanisms}
\label{SS:discussion_evol}
Having established the relations between SFR, $M_{\rm mol}$, and $M_*$ across quenching stages, and the relative importance of SFE and $f_{\rm mol}$, it is now natural to discuss the dominant causes responsible for the star formation quenching. Here, we are implicitly assuming that a sort of evolutionary path from one stage to another exists and that follows the average distance of the galaxies from the SFMS for a given group. Considering the decrease in the relative fraction of star-forming spaxels in favour of retired spaxels (discussed in Section~\ref{SSS:sample_stats_questna}), this might be the case. 

To move from fully star forming to galaxies quenched in their centre (QnR and cQ groups), galaxies need to reduce their molecular gas mass per stellar mass. In the literature, AGN feedback is often considered as the fundamental process that removes gas from galaxies and keeps it ionised within the halo (see \citealt{piotrowska2022,goubert2024}). In principle, we have excluded active galaxies from the analyses presented in this paper, allowing us to rule out AGN-driven gas depletion as the primary mechanism for quenching in our sample. However, active galaxies were identified through BPT diagrams and narrow optical emission lines, which are subject to classification biases compared to more robust methods like X-ray emission-based selection \citep{bickley2024}. Consequently, some AGN-host galaxies might still be lurking in our subsample. Nevertheless, considering that AGN-hosting galaxies make up a small fraction of the local Universe population (up to 10\%, as we measured, consistent with \citealt{lacerda2020,kalinova2021,osorio-clavijo2023}), it is plausible that nuclear activity is a consequence rather than a direct cause of quenching. This aligns with our finding that nuclearly active galaxies in our sample tend to be brighter in CO than their non-active counterparts, even at the same quenching stage.

Such observations can be understood within a framework that incorporates time-scale dependencies between AGN activity and quenching. Specifically, quiescence may emerge as a long-term consequence of AGN heating preventing gas cooling and accretion from the circumgalactic medium (CGM), effectively starving galaxies of the fuel required for sustained star formation (e.g. \citealt{fukugita2004}, \citealt{fabian2012}, \citealt{bluck2014}, \citealt{bluck2023}). While instantaneous feedback from AGN may trigger initial quenching \citep{terrazas2020,zinger2020}, the lack of sustained CGM heating would allow gas to cool and condense back into the galaxy, reigniting star formation. This highlights the complex interplay of feedback mechanisms operating over varying time-scales, suggesting that the observed cessation of star formation is not solely a direct and immediate result of AGN activity but also a cumulative outcome of long-term energy release and preventative feedback processes. The detailed analysis of these effects, however, is outside the scope of the paper and will be explored elsewhere \citep{bazzi2025}. 

The effect of environments in which galaxies live can certainly play a role in removing the gas from galaxies. Nevertheless, \cite{garay-solis2023}, using the sample of \cite{colombo2020}, showed that the gas fraction in the centre of interacting galaxies is enhanced compared to the isolated counterparts and that the reduced SFR might be due to the highly turbulent nature of this gas. The merger stages classification by \cite{garay-solis2023} also indicated that a significant number of interacting systems is present in our sample.

The interpretation that lower $f_{\rm mol}$ values in cQ and QnR galaxies compared to SF ones indicate a reduced molecular gas supply for star formation might be misleading. Indeed, SF, cQ, and QnR galaxies exhibit similar average $M_{\rm mol}$ globally. Instead, the key difference appears to lie in the stellar mass, which is approximately 0.5 dex higher (on average) in cQ and QnR galaxies compared to SF ones (see Fig.~\ref{F:flaring}). This suggests the existence of a specific `mass threshold' \citep[e.g.][]{baldry2006,woo2013}, beyond which galaxies may trigger quenching mechanisms such as AGN feedback \citep[e.g.][]{weinmann2006, Peng2010} and morphological transformations \citep[e.g.][]{Kauffmann2003} that induce large-scale dynamical processes leading to gaseous disc stabilisation. The commonly cited threshold is $10^{10.5}\,M_{\odot}$, which roughly corresponds to the mass above which most cQ and QnR galaxies are observed.

Indeed, large-scale dynamics are often invoked as plausible mechanisms that bring galaxies to quiescence. For example, the results by \cite{kalinova2022} established that a direct link between the galaxy quenching stage and circular velocity curve shape and amplitude exists. By itself, the circular velocity curve is a simplification of the galactic potential that averages several dynamical effects. The theoretical work of Gensior et al. (2020) suggested that a dynamical suppression of the star formation can lead to quenching. In this scenario, central stellar spheroids (such as bulges) increase the turbulence of the gas (therefore, the stable mass of the molecular clouds), reducing the gas fragmentation and the self-regulation of star formation by stellar feedback. As observed by \citetalias{kalinova2021} (see their Fig.~9), cQ galaxies show some of the highest bulge-over-disc ratios between the quenching stages dominated by star formation. The presence of a bulge can be the cause that initiates the quenching in this group. Instead, in \citetalias{kalinova2021}, the authors found that QnR galaxies show the highest fractions of bars in their sample. It has been proven that the presence of bars in the galaxies can have a significant effect on the star formation quenching. Bars can drive secular evolution by redistributing angular momentum, funneling gas towards the central regions, and inducing starbursts followed by gas depletion, leading to a decline in the overall star formation activity \citep[e.g.][]{spinoso2017}. For instance, \cite{masters2010} and \cite{masters2012} found that barred galaxies tend to have lower star formation rates compared to their unbarred counterparts, particularly in the green valley population. \cite{cheung2013} showed that strong bars are associated with a suppression of star formation in the central regions of galaxies, likely due to the rapid depletion of molecular gas and the stabilisation of gaseous discs against collapse. \cite{khoperskov2018} found that barred galaxies exhibit signs of quenched star formation, particularly in the inner regions, suggesting that bars contribute to the suppression of star formation by affecting the distribution and dynamics of gas within galaxies. Nevertheless, \cite{hogarth2024} demonstrated that the strength of the bar inflows rather than the mere presence of bars in the galaxies to influence the star formation rate.
Given also that the cQ and QnR groups presented almost identical fractions of star-forming and retired spaxels, the two patterns might represent two different evolution channels (driven by bulge and bar-associated dynamics, respectively) that bring the galaxies to the green valley and, eventually, full retirement. 

However, while these quenching stages exhibit prominently suppressed star formation in their central regions, their global (average) star formation efficiency remains comparable to that of SF galaxies. This suggests that dynamical suppression may play a secondary role, becoming significant (or even dominant) only after a substantial amount of gas has already been removed from the galaxy. \cite{gensior2021} found that this critical gas fraction aligns with the average molecular gas fraction observed in the MX group. Notably, the MX stage marks the point where a noticeable decline in star formation efficiency compared to SF galaxies emerges. Since dynamical suppression requires the presence of a significant molecular gas reservoir that remains inefficient at forming stars, it naturally leads to reduced star formation efficiency \citep{colombo2018,ellison2020}. This mechanism primarily operates through shear-driven gas stabilisation. Indeed, rotation curve shear might be the parameter that sets the link between the quenching stage and circular velocity curve type \citep{kalinova2022}. Shear can be induced in other galactic regions where reduced SFE are observed such as galactic centres (\citealt{kruijssen_longmore14}, where also gas velocity dispersions are higher compared to the discs, e.g. \citealt{colombo14a,leroy2016,sun2018,duarte-cabral2021}), bars \citep[e.g. ][]{athanassoula92,hogarth2024,scaloni2024}, but even within strong spiral structure \citep[e.g][]{meidt13,dobbs_baba2014}, and in locations where gas flows are observed \citep[e.g. ][]{meidt2018}.

Large-scale dynamical processes and their impact on quenching may be even more pronounced in the retired stages, nR and fR. Galaxies in these evolutionary phases exhibit prominent spheroids and the highest shear values among the quenching stages \citep{kalinova2021,kalinova2022}. Furthermore, these galaxies are the most gas-poor in the sample, which could further amplify the role of dynamical suppression in halting star formation.

\section{Summary}\label{S:summary}
In this paper, we have investigated the relationships between SFR, $M_*$, and $M_{\rm mol}$ and their ratios (sSFR, SFE, $f_{\rm mol}$) across galaxies at different evolutionary phases. To do so, we used \iedge\ a homogenised database that contains information on the integrated stellar measurements, optical emission lines, and CO lines drawn from CALIFA and EDGE observations (with APEX, CARMA, and ACA) of 643 galaxies. Our most important results are summarised as follows:

\begin{itemize}

    \item The quenching stage and nuclear activity of the galaxies within the \iedge\ were classified using the \questna\ scheme by considering the pattern defined by regions in the $W_{\rm H\alpha}$ maps and BPT diagrams. We found that 60\% of the galaxies in the database are largely dominated by star formation (e.g. SF, QnR, and cQ stages), 17\% of the galaxies are transitionary objects with undefined $W_{\rm H\alpha}$ morphology, and 23\% of the targets are largely retired (e.g. the nR and fR stages). Less than 10\% of the galaxies in the sample (not included in the analyses of this paper) are nuclearly active.

    \item While the star formation efficiency is (on average) constant across the SF, QnR, and cQ stages and largely decreases from the MX to the fR classes, the fraction of molecular gas constantly decreases along the quenching stages sequence.

    \item The global variation of $M_{\rm mol}$ and $M_*$ and their related quantities (such as $f_{\rm mol}$) with respect to the SF class reflect the variation in the inner galactic regions. When considering SFR and related quantities (sSFR and SFE), we instead observed that these variations are more significant in the inner galactic regions than globally. This suggests that while galaxies are brought to the green valley by a reduction of the availability of molecular gas, the inside-out quenching from the green valley is mostly driven by a decrease in SFE, which is more prominent in the inner galaxies than in their outskirts.

    \item By examining the scaling relations between SFR, $M_*$, and $M_{\rm mol}$, we observed that when decreasing, the amount of molecular gas becomes less correlated with the stellar mass across the quenching stages. The relations between SFR and $M_{\rm mol}$ appear increasingly steeper and have a greater amount of scatter. This indicates that star formation in galaxies is still largely dependent on the amount of available molecular gas, but passive systems are unable to convert this gas efficiently into stars. Additionally, passive systems are unable to sustain large amounts of molecular gas.

    \item A three-dimensional relation between SFR, $M_*$, and $M_{\rm mol}$ appears to be defined only for purely star-forming galaxies (SF group). We speculate that this might be due to lower stellar feedback (and interstellar pressure), which reduces the self-regulation of the star formation in quenching galaxies, or to the bias in our CALIFA sub-sample, where quenched and retired low-mass galaxies are under-represented. 

\end{itemize}

As AGN-host galaxies were excluded from our sample, we speculate that the significant reduction of SFE (especially in the centre of the galaxies) that drives the progression of the quenching is due to a combination of large-scale dynamics and, in particular, the action of shear (in different forms) that essentially decreases molecular gas fragmentation or increases the stable mass of molecular clouds, thus preventing efficient star formation. However, to extensively test this, integrated quantities are not enough, and progress can only be made by observing the molecular gas at cloud scale (with, e.g. ALMA or NOEMA) for a similar-sized sample of galaxies at different stages of their evolution.

\begin{acknowledgements}
The authors sincerely thank the anonymous referee for the important insights that significantly enhanced the clarity of the paper. DC thanks Mallory Thorp for the fruitful discussions about galaxy evolution. DC, ZB, and FB gratefully acknowledge the Collaborative Research
Center 1601 (SFB 1601 sub-project B3) funded by the Deutsche Forschungsgemeinschaft (DFG, German Research Foundation) –
500700252. DC and AW acknowledge support by the \emph{Deut\-sche For\-schungs\-ge\-mein\-schaft, DFG\/} project number SFB956-A3. SFS thanks the PAPIIT-DGAPA AG100622 project and CONACYT grant CF19-39578. This work was supported by UNAM PASPA – DGAPA. ER acknowledges the support of the Natural Sciences and Engineering Research Council of Canada (NSERC), funding reference number RGPIN-2017-03987. ADB and TW acknowledge support from the National Science Foundation (NSF) through the collaborative research award AST-1615960. KDF acknowledges support from NSF grant AAG 23-07440. TW acknowledges support from the NSF through grant AST-1616199. JBB acknowledges support from the grant IA-100420 (PAPIIT-DGAPA, UNAM). JBB acknowledges funding from the grant IA-101522 (DGAPA-PAPIIT, UNAM) and support from the DGAPA-PASPA 2025 fellowship (UNAM). VV acknowledges support from the ALMA-ANID Postdoctoral Fellowship under the award ASTRO21-0062. This research made use of Astropy,\footnote{http://www.astropy.org} a community-developed core Python package for Astronomy \citep{astropy2013, astropy2018}; matplotlib \citep{matplotlib2007}; numpy and scipy \citep{scipy2020}. Support for CARMA construction was derived from the Gordon and Betty Moore Foundation, the Eileen and Kenneth Norris Foundation, the Caltech Associates, the states of California, Illinois, and Maryland, and the NSF. Funding for CARMA development and operations was supported by NSF and the CARMA partner universities.
\end{acknowledgements}

\footnotesize{
\bibliographystyle{aa}
\bibliography{cold}

\begin{appendix}

\section{Automatic \questna\ classification}
\label{A:auto_class}

Following the prescriptions of \citetalias{kalinova2021}]{kalinova2021}, we have created an automatic routine that helps to benchmark the by-eye \questna\ classification results. In the following, we describe it in detail. 

In \citetalias{kalinova2021}]{kalinova2021}, the quenching stages are established considering only patterns in the $W_{\rm H\alpha}$ maps. Therefore, following this, we analyse only \wha\ values where $W_{\rm H\alpha}/\delta W_{\rm H\alpha}>1$, where $\delta W_{\rm H\alpha}$ indicates the uncertainties of the $W_{\rm H\alpha}$ value from PIPE3D.  

We firstly distinguish three groups of quenching stages, the retired group, i.e. galaxies having median \wha\ ($\left<W_{\rm H\alpha}\right>$) across the \wha\ map $<3$\,\AA; the star-forming group, where median $\left<W_{\rm H\alpha}\right> >6$\,\AA; and the mixed group, where the median is within these limits, e.g. $3<\left<W_{\rm H\alpha}\right><6$\,\AA.

Within the retired group, we classify galaxies as part of the nR stage if they have at least one region where all constituting pixels must have $W_{\rm H\alpha}>3$\,\AA. The region needs to be at least 3 pixels large and contained within 2\,$R_{\rm e}$. If this condition is not satisfied, the galaxy is classified as fR.

Within the star-forming group, we first search for the QnR galaxies. According to the general definition from \citetalias{kalinova2021}]{kalinova2021}, medians \wha\ across the map need to be as follows: $\left<W_{\rm H\alpha}\right>_{R/R_{\rm e} < 0.2} >3$\,\AA, $\left<W_{\rm H\alpha}\right>_{0.2 < R/R_{\rm e} < 0.5} < 3$\,\AA, and $\left<W_{\rm H\alpha}\right>_{0.5 < R/R_{\rm e} < 2} < 6$\,\AA. Additionally, we need to make sure that the galaxy indeed has a central quenched ring, which might appear as discontinuous for highly inclined objects. To do so, we searched for a closed circular structure with $W_{\rm H\alpha} >3$\,\AA\ or $W_{\rm H\alpha} >6$\,\AA\ within $R/R_{\rm e} < 0.2$. The circular structure must consist of at least 3 pixels. If those conditions are not satisfied, we try to classify the galaxy as cQ. For this, the galaxy must have a circular structure within $R/R_{\rm e} < 0.5$ consisting of at least 3 pixels with $W_{\rm H\alpha} < 3$\,\AA. Additionally, the galaxy must have $\left<W_{\rm H\alpha}\right>_{0.5 < R/R_{\rm e} < 2} >3$\,\AA. All objects of the {star-forming} group whose properties do not satisfy these conditions are collected within the SF quenching stage.

Additionally, we assume 3 quality flags to assess the reliability of the automatic attribution of the quenching stages. We attribute a quality flag equal to two to all objects that have been classified to a group or another due a borderline global $\left<W_{\rm H\alpha}\right>$ value, i.e. star-forming group objects with $6 < \left<W_{\rm H\alpha}\right> < 7$\,\AA\, retired group objects with $2 < \left<W_{\rm H\alpha}\right> < 3$\,\AA\, and mixed objects with $3.5 < \left<W_{\rm H\alpha}\right> < 5.5$\,\AA{}. A quality flag value equal to three deals with possible misclassifications due to low S/N pixels. As we mentioned, we considered all pixels in the \wha\ maps that have $W_{\rm H\alpha}/\delta W_{\rm H\alpha}>1$. Nevertheless, a fully reliable classification should give consistent results when only high S/N pixels are considered. We consider a quality flag equal to 3 for galaxies where $\left<W_{\rm H\alpha}\right> > 6 \mathrm{(3)}$\,\AA\ if all pixels with $W_{\rm H\alpha}/\delta W_{\rm H\alpha}>1$ are considered, but $\left<W_{\rm H\alpha}\right> < 6 \mathrm{(3)}$\,\AA\ where only pixels with $W_{\rm H\alpha}/\delta W_{\rm H\alpha}>3$ are used to calculate the median. In other words, a quality flag equal to three indicates that a galaxy might have been classified in the mixed group (instead of the star-forming or retired group) if the classification is performed considering only high S/N pixels, for example. Galaxies that do not satisfy these conditions have the most reliable classification and assume a quality flag equal to one.

Parallelly, we automatically establish the nuclear activity of the galaxies. In essence, we use BPT diagram models together with \wha\ maps to classify the galaxies into nonA, wAGN or sAGN. We first searched for a region within $R/R_{\rm e} < 0.2$ that populated the Seyfert region of the BPT diagram. This region must consist of at least three pixels and be present in at least two of the three BPT diagrams. Additionally, if at least three pixels in this region have $W_{\rm H\alpha}>6$\,\AA{}, the galaxy is classified as sAGN. However, if the region has at least three pixels with $W_{\rm H\alpha}>3$\,\AA{} but less than three pixels with $W_{\rm H\alpha}>6$\,\AA{}, the galaxy is classified as wAGN. In every other case, the galaxy is classified as nonA.

As in the case of the quenching stage, quality flags are also used here to assess the reliability of the automatic nuclear activity classification. If the signal of a strong or weak AGN is shown only in 2 over 3 BPT diagrams, the galaxies are classified as sAGN or wAGN, respectively, but with a quality flag equal to `U' (that stands for `unsure'). If the signal of an AGN is shown by all 3 diagrams, the quality flag is set to `S' (or `secure' classification). Additionally, if one diagram indicates the presence of a strong AGN, another indicates the presence of a weak AGN, and the third one indicates that the galaxy is not active, the galaxy is classified as wAGN with a quality flag equal to ``U''. If only one diagram shows the presence of an AGN, the galaxy is classified as nonA, with a quality flag set to ``U''. In the case that all diagrams do not show the presence of an AGN, the galaxy is classified as nonA, with a ``S'' quality flag. 

Given our sample, the automatic classification agrees with the visual classification of the quenching stage for 597/643 galaxies, i.e. $\sim93$\% of the cases. For 11 mismatching galaxies, the automatic classification gives a quality flag equal to 1, and 35 galaxies show a quality flag equal to 2. We explore the \wha\ pattern case by case regarding the mismatching galaxies, and we assume the visual classification for 46 objects.

The automatic classification agrees with the visual classification for 630/643 galaxies, regarding the nuclear activity (i.e. roughly 98\% of the sample). Only 10 of these galaxies have an ``unsure'' quality flag attributed by the automatic classification. 
After careful inspection of all these diverging cases, we attributed the result to the visual classification of these 13 galaxies.

\end{appendix}

\end{document}